\documentclass[aps,showpacs,10pt]{revtex4-1}
\usepackage{amsmath,amssymb,mathrsfs}
\usepackage{graphicx}
\usepackage{epsfig}
\usepackage[usenames]{color}
\usepackage{times}
\usepackage{ulem}
\usepackage{cancel}
\usepackage[colorlinks=true,urlcolor=blue,citecolor=blue]{hyperref}

\newcommand{\beq}{\begin{equation}}
\newcommand{\eeq}{\end{equation}}

\begin{document} 

\title{Long-range interactions from \texorpdfstring{$U\left(1\right)$}{Lg} 
gauge fields via dimensional mismatch}

\author{Joao C. Pinto Barros}
\affiliation{SISSA and INFN, Sezione di Trieste, Via Bonomea 265, I-34136 Trieste, Italy}

\author{Marcello Dalmonte}
\affiliation{Abdus Salam International Center for Theoretical Physics, Strada Costiera 11, Trieste, Italy}

\author{Andrea Trombettoni}
\affiliation{CNR-IOM DEMOCRITOS Simulation Center, Via Bonomea 265, I-34136 Trieste, Italy}
\affiliation{SISSA and INFN, Sezione di Trieste, Via Bonomea 265, I-34136 Trieste, Italy}

\begin{abstract}
We show how certain long-range models of interacting fermions 
in $d+1$ dimensions are equivalent to $U\left(1\right)$ gauge theories in 
$D+1$ dimensions, with the dimension $D$ in which gauge fields are defined 
larger than the dimension $d$ of the fermionic theory to be simulated. 
For $d=1$ it is possible to obtain an exact mapping, providing an 
expression of the fermionic interaction potential in terms of 
half-integer powers of the Laplacian. An analogous mapping 
can be applied to the kinetic term of the bosonized theory. 
A diagrammatic representation of
the theories obtained by dimensional mismatch is presented, and consequences and applications of the established duality are discussed. 
Finally, by using a perturbative approach, we address 
the canonical quantization of fermionic theories presenting non-locality 
in the interaction term to 
construct the Hamiltonians for the effective theories found by dimensional 
reduction. We conclude by showing that one can engineer 
the gauge fields and the dimensional mismatch in order 
to obtain long-range effective Hamiltonians with $1/r$ potentials.
\end{abstract}

\maketitle

\section{Introduction}
\label{intro}

In the last decade, there has been significant progress in the realization of synthetic quantum matter \cite{Bloch2008} via the control and manipulation of AMO systems  
such as trapped ions, Rydberg atoms, and quantum gases of magnetic atoms and 
polar molecules \cite{Lahaye2009,Blatt2012,Saffman2010,Blatt2012,Ritsch2013}. 
Such advancements made possible the implementation  
of a variety of quantum long-range (LR) models 
\cite{Britton2012,Schauss2012,Aikawa2012,Lu2012,Yan2013,Islam2013,Richerme2014,Jurcevic2014,Douglas2015,Schempp2015,Landig2015,Landig2015a}. 
Recent highlights in this directions include the physical 
realization of Ising and $XY$ quantum spin chains 
with tunable LR interactions 
with ions in a Penning trap 
\cite{Britton2012}, neutral atoms in 
a cavity \cite{Douglas2015,Landig2015,Landig2015a},
trapped ions \cite{Islam2013,Richerme2014,Jurcevic2014}, 
and Rydberg atoms \cite{Schauss2012}. 
The resulting interactions decay algebraically with 
the distance $r$, and the decay exponent can be experimentally tuned 
\cite{Islam2013,Richerme2014,Jurcevic2014}. As an example of the many 
and promising possibilities opened by the control of LR interactions 
in spin chains we may refer to the recent emulation 
of the one-dimensional ($1D$) version of quantum electrodynamics, the Schwinger model \cite{Martinez2016}.

Such experimental progresses found a counterpart in  
the intense theoretical activity on the properties of quantum LR systems 
\cite{Laflorencie2005,Hastings2005,Koffel2012,Schachenmayer2013,Eisert2013,Gong2014,Damanik2014,Vodola2014,Ares2015,Gori2015,Viyuela2015,Gong2016,Lepori2016,Lepori2016a,Maghrebi2016,Santos2016,Fey2016,Gong2016a,Kovacs2016,Humeniuk2016,Bermudez2016,Lepori2016b,Defenu2017}. Among others, we mention 
the study of the effect of non-local interactions 
on the dynamics of excitations \cite{Hauke2013,Foss-Feig2015,Rajabpour2015,Cevolani2015,Kuwahara2016,VanRegemortel2016,Buyskikh2016,Bertini2015} 
and on the equilibrium properties and phase diagram 
\cite{Brezin2014,Angelini2014,Defenu2015,Behan2017,Behan2017a}. This recent 
activity aims in perspective also to provide a quantum counterpart 
of the well established results for classical LR systems \cite{Campa2014}.

Usually, power-law interactions are written as $V(r)\propto r^{-d-\sigma}$,
where $r$ is the distance between the particles or spins, $d$ is the physical space dimension, and 
$\sigma$ is the decay exponent. 
For $\sigma<0$, the internal energy of 
the system typically diverge
in the thermodynamic limit, calling for a redefinition of the 
interaction strength \cite{Kac1963}. For fast enough decaying interactions, $\sigma > 0$, 
the system is additive and thermodynamics is well defined. 
For $\sigma$ belonging to a certain range, the system may present
a phase transition with spontaneous symmetry breaking at
a critical temperature $T_{c}>0$ \cite{Sak1973}.  
For $2\sigma \le d$ one has mean field properties. However one can define 
a critical value $\sigma^{*}$ such that for $\sigma > \sigma^{*}$ 
the system has the same critical behavior of its short-range (SR) analogue.
In the intermediate region $d/2 < \sigma \le \sigma^{*}$ 
the LR interactions are found to be relevant and the determination 
$\sigma^{*}$ has been the subject of a long-lasting theoretical activity 
\cite{Brezin2014,Angelini2014,Defenu2015,Behan2017,Behan2017a}. 

The main point of this analysis is that increasing $\sigma$ 
one is effectively passing from the upper critical 
dimension to the lower critical one. Therefore to each 
$\sigma$ there is, at least effectively, an effective dimensionality 
$d_{\mathrm{eff}}$ so that 
the LR system is equivalent to a SR model living in $d_{\mathrm{eff}}$. 
In concrete, this in principle allows one to apply well 
established results for local many-body systems, such as the 
Mermin-Wagner-Hohenberg theorem~\cite{mermin1966absence,hohenberg1967existence} and Lieb-Robinson bounds~\cite{vershynina2013lieb} on the 
propagation of quantum correlations, to non-local models. However, 
it has been challenging to cast this intuitive picture into a 
rigorous theoretical framework, mostly due to the complexity of the 
mapping directly at the operator (i.e., Hamiltonian) level.

Here, we propose a new approach to the problem, 
based on an exact mapping between LR interacting systems in 
$d$ dimensions, and particles coupled to dynamical gauge fields in a higher 
dimension $D>d$. The central aspect of our work is 
the consideration that, when fermionic degrees of freedom are confined 
to move on a reduced dimensionality with respect to the gauge fields, 
the latter can mediate different types of interactions thanks 
to the extra dimension(s) available, effectively providing 
a knob to tune the interparticle potential. 
Since the initial theory is fully local in $D$ spatial dimensions, 
this mapping allows one to exploit the full predictive power of general 
results for local field theories to the non-local one in $D$ dimensions. 
As an illustrative case sample, we exploit this strategy to discuss fermionic 
(and spin) models in $D=1$, and we show that the latter systems in the presence of Coulomb interactions can be exactly mapped to Abelian gauge theories 
living in higher dimensions.

Our work takes inspiration from graphene experiments \cite{CastroNeto}, 
where the system 
is two-dimensional, but the electro-magnetic field acting on it 
is living in three dimensions. Specifically, in such settings,
the electrons are confined in a $2D$ (i.e., $2+1$) plane 
while interacting with the electromagnetic field that lives in the full 
$3D$ space (i.e., $3+1$). The formalism of Pseudo QED, introduced in 
\cite{Marino93}, provides a full dynamic way, from first principles, 
to deal with specific problem. 
For a system confined in a two-dimensional space, with $\left(x,y,z=0\right)$
coordinates, this is done by taking the fermion kinetic part of the 
two-dimensional space and the electromagnetic kinetic term in 
three-dimensional space. The two fields are then coupled 
through the standard minimal coupling with the additional requirement 
that no fermionic current exists or flows outside $z=0$. 
In standard QED, one writes the $4$-current of the fermions in $3+1$ 
dimensions in the form $j_{4}^{\mu}=\bar{\psi}\gamma_{\mu}\psi$. 
Here, however, there is a dimensional mismatch and the gamma 
matrices indices do not run through the same set of numbers. 
This is overcome by considering
\begin{equation}
j_{4}^{\mu}\left(\tau,x,y\right)=\left\{ \begin{array}{ll}
j_{3}^{\mu}\left(\tau,x,y\right)\delta\left(z\right)\ \mathrm{if}\ \mu=0,1,2\\
0 \hspace{65pt} \mathrm{otherwise}
\end{array}\right.\label{eq:from3to2}
\end{equation}
where $j_{3}^{\alpha}=\bar{\psi}\gamma_{\alpha} \psi$ is the $3$-current of the fermions in $2+1$ dimensions. Eq. (\ref{eq:from3to2}) 
states that no fermionic current exists or flows outside the plane.
By integrating out the gauge field and applying the above condition 
(\ref{eq:from3to2}) the
resulting Lagrangian consists on an effective $2+1$ dimensional
Lagrangian containing a LR interaction 
\cite{Marino93,Marino14,Kotikov2014,Marino2015,Nascimento2015,Alves2017,Menezes2017}. This LR term is 
fundamentally different from the LR interaction obtained when 
the original electromagnetic field is 
also living in the same dimensionality as the fermions ($2+1$) and is at the 
basis of several peculiar properties of Pseudo QED, such as the 
dynamical generation of a mass term \cite{Marino2015,Alves2017}. 
The dynamical chiral symmetry breaking in reduced QED theories was 
studied as well~\cite{Miransky2001}, and the procedure of dimensional reduction 
was applied to the edge modes of two-dimensional topological insulators 
\cite{Menezes2016}.

In the present work we start by generalizing the Pseudo QED 
construction to general dimensions. The dimensional reduction giving 
a LR term in $2+1$ dimension starting from a theory in which the electromagnetic 
(fermionic) field lives in $3$ ($2$) spatial dimensions can be indeed seen from 
the opposite point of view, i.e. {\it i)} 
determine what is the gauge field living 
in higher dimensions giving rise to a target LR interaction to be implemented 
via the dimensional mismatch,  
and {\it ii)} explore what kind of LR interactions can be realized. We then consider the general scenario where the electromagnetic field 
and general $U(1)$ gauge fields 
live in $D+1$ dimensions and the fermions in $d+1$ with $D\geq d$. 
The case $D=d$ is of course trivial in the sense that it corresponds to QED$d$, 
even though it is useful to keep in mind that the known results for 
this case should be recovered whenever we take $D=d$. 

Lifting the restriction $D=d$ opens the door to new kind of systems with novel 
LR interactions but which are derived, from first principles, 
from a local gauge theory. Such formulations may not only be useful from 
a direct application point of view, as in the graphene example, 
but may also serve as a tool to characterize LR systems. 
In fact, by being able to map a LR interacting system to a 
local gauge theory one may be able to use tools otherwise unaccessible 
in LR systems like the Mermin-Wagner theorem or Lieb-Robinson bounds. 

Another motivation behind the present work is that in literature it has been 
discussed the possibility to have tunable interactions with cold atoms, 
and in particular $1/r$ interactions \cite{ODell2000}. Despite the fact that, for trapped 
ions, interactions of Ising-type 
can be made to decay algebraically with the distance $r$ 
with an adjustable exponent (usually in the range $\lesssim 3$), so far 
no experiments have been performed for a Bose or Fermi gas with an effective 
$1/r$ interaction also in lower dimension 
and proposals in this direction are certainly desirable.

With these various motivations 
in mind, one may consider the general scenario where there are 
$N_f$ fermionic flavors $\left\{\psi_a\right\}_{a=0,...,N_f}$ living in 
$d+1$ dimensions coupled to $N_g$ gauge fields $\left\{A_{\mu}^{b}\right\}_{b=0,...,N_g}$ each one living in $D_b+1$ dimensions (where $D_b\geq d, \forall b$) 
with coupling parameters $e_{ab}$. Furthermore, 
with the advent of quantum simulations of gauge theories~\cite{wiese13,zohar15,dalmonte16} 
in which gauge symmetries are engineered in the laboratory, 
these systems can open the possibility to explore new kind of phenomena 
or provide humbler toy models to more complicated systems. 
A first demonstration of a quantum simulation of a gauge theory was put 
forward in \cite{Martinez2016}. In a near future it is expected that more 
complex theories may be achieved from which the above described 
scenario constitutes a particular example. Indeed, the fact 
that the fermions are restricted to a lower dimension
constitutes a simplification when compared to the case where all particles
live in the higher dimensional system. For reviews on the topic of quantum 
simulation of gauge theories see e.g. \cite{wiese13,zohar15,dalmonte16}. 

In this work, our attention is focused on the effective interaction
between fermionic degrees of freedom. The gauge fields will be taken in three possible dimensions $D=1,2,3$, minimally 
coupled with the fermions. This structure, of dimensionality and type of coupling, poses restrictions on the type of interactions 
obtained in the effective fermionic theory. Still, in the case of $d=1$, 
fermionic degrees of freedom can also be integrated out paving the way 
to more general interactions for a given fermionic flavor 
(left out of the integration). Further generalizations 
are possible which may incorporate Higgs fields, interaction between 
gauge fields ($F_{\mu\nu}^aF_{\mu\nu}^b$) or other symmetries besides 
$U\left(1\right)$. Such generalizations are beyond the scope of this paper 
where we target in detail the construction of general interacting potential 
between fermions and kinetic terms of bosonic theories in $d=1$. 
The mapping is achieved by bosonizing all fermionic degrees of freedom in the last case, and all but one on the first. 
We consider also the possibility of a four fermion interaction in the form 
of a current-current term $j_\mu j_\mu$ where $j_\mu=\bar{\psi}\gamma_\mu \psi$.

The paper is organized as follows: In Section \ref{sec:fromD_to_d} we establish 
the general formalism for the construction of the effective 
fermionic theory in the lower dimensionality ($d+1$) and its respective 
relation with the gauge theory, in the same dimension, with 
a modified gauge kinetic term. In Section \ref{sec:Exploring_construction} 
we consider the special case $D=2$ and $d=1$ where we use bosonization. 
We show how it is possible to consider several fermionic flavors and 
gauge fields in such a way that, integrating all of the gauge fields 
and bosonizing all the fermionic flavors it is possible to construct a 
general bosonic kinetic term, on the Lagrangian. of the form 
$\phi f\left(-\partial_{\mu}^{2}\right)\phi$. Here $\phi$ is the bosonic field and $f$ is a fucntion which can be seen as
an expansion in half-integer exponents
(in powers of $\alpha$ with $2\alpha \in \mathbb{Z}$). When  $f$ is the identity function we have the standard kinetic term for a bosonic field. We have freedom 
on engineering the coefficients of these expansions by modifying 
the coupling of the initial theory. 
By following the same process bosonizing all fermionic flavors but one, 
it is also possible to construct a similar kind of interaction between fermions $j_\mu V\left(-\partial_{\nu}^{2}\right)j_\mu$ again with $V$ admitting  a series expansion on intenger and half-integer powers of the Laplacian and $j_\mu$ the fermionic current. In Section \ref{exp_implementation} 
we provide an overview on how these kind of models can naturally fit in the 
class of proposals  of experimental realization of quantum simulations of 
gauge theories available on the literature. In Section \ref{LRH} we
deal with the canonical quantization and the construction of the 
Hamiltonians for the models obtained by dimensional reduction, and 
show how to obtain non-relativistic fermions 
interacting via an $1/r$ potential. Our conclusions are presented 
in Section \ref{conclusions}, while more technical material is in the 
Appendices.

\section{Dimensional reduction}
\label{sec:fromD_to_d}

We start by reviewing the formalism of Pseudo QED, referring to 
electrons confined in a plane and interacting with an 
electromagnetic field defined in the $3D$.
To make this statement explicit, one takes a matter 
Lagrangian $\cal{L_M}$ coupled with a gauge field (in Euclidean time):
\begin{equation}
{\cal L}={\cal L}_M-ie j^{\mu}_{4}A_{\mu}+\frac{1}{4}F_{\mu\nu}^{2},
\label{eq:PQED}
\end{equation}
where $j_{4}^{\mu}=\bar{\psi}\gamma_{\mu}\psi$. The $4$-current $j_{4}$ 
and the $3$-current defined $j_{3}$ defined in the space 
$(x, y, z = 0)$ are related by Eq. (\ref{eq:from3to2}). One has then 
to integrate the gauge field and apply the condition  (\ref{eq:from3to2}), 
obtaining a $2+1$ Lagrangian \cite{Marino93}.
It is possible to reformulate such theory and re-write it in terms
of a three-dimensional gauge field and re-storing locality in the
fermionic part of the Lagrangian. The price to pay, 
which is actually a resource for the purposes of our paper, 
consists in transferring the long range term to 
the kinetic part of this new gauge field. Nonetheless, $U\left(1\right)$
gauge invariance is retained. 

In direct generalization of (\ref{eq:PQED}), we consider 
the Lagrangian
\begin{equation}
{\cal L}={\cal L}^{d+1}_M-ie j^{\mu}_{D+1}A_{\mu}+\frac{1}{4}F_{\mu\nu}^{2}+{\cal L}_{GF}.\label{eq:QEDLagrangian}
\end{equation}
The term ${\cal L}_{GF}$ corresponds to the Fadeev-Popov gauge fixing
term. It is explicitly given 
by ${\cal L}_{GF}=\frac{1}{2\xi}\left(\partial_{\mu}A_{\mu}\right)^{2}$, 
where different choices of $\xi$ correspond to different gauges. 
Here we adopt the Feynman gauge where $\xi=1$, 
resulting in a propagator $G_{\mu\nu}=\frac{1}{-\partial^{2}}\delta_{\mu\nu}$.
The fermions are defined in the lower dimensionality $d+1$, 
which is made explicit in the matter Lagrangian ${\cal L}^{d+1}_M$ . 
The gauge field lives in higher dimensionality $D+1$. 
The $D+1$ current is taken explicitly to be:
\begin{equation}
j_{D+1}^{\mu}\left(x^{\alpha}\right)=\left\{ \begin{array}{ll}
j_{d+1}^{\mu}\left(x{}_{0},\ldots,x_{d}\right)\delta\left(x{}_{d+1}\right)\ldots\delta\left(x_{D}\right)\ \mathrm{if}\ \mu=0,\ldots,d\\
0\hspace{149pt} \mathrm{otherwise}
\end{array}\right.\label{eq:fromDtod}
\end{equation}
  
From this point one can integrate out the gauge degrees of freedom 
obtaining then a non-local interaction between the fermions of the form 
$$\frac{e^{2}}{2}\int j_{D}^{\mu}\left(z\right)\left[\left(-\partial^{2}\right)^{-1}\right]_{zz^{\prime}}j_{D}^{\mu}\left(z^{\prime}\right)d^{D+1}z^{\prime}.$$
We shall denote this long range potential by $G_{D}\left(z-z^{\prime}\right)$
exhibiting explicitly the dependence on the dimension where the gauge fields lived. The nature 
of these interactions is encoded in the higher dimension $D+1$ and it is
not dependent on the dimensionality where the fermions live
which did not enter yet. We observe that 
the fact that integrating degrees of freedom one obtains 
LR terms (and possible multi-body interactions) is ubiquitous in 
RG treatments of models, where typically one takes a model and integrate 
over a sub-class of the original degrees of freedom remaining in the same 
dimensional space (see, for example, \cite{cardy1996}). The difference with the models  
considered here is then that one performs a dimensional 
reduction while making the integration of gauge degrees of freedom. 

An explicit expression for $G_{D}$ can be
obtained as 
\begin{equation}
G_{D}\left(z-z^{\prime}\right)=\int\frac{d^{D+1}k}{\left(2\pi\right)^{D+1}}\frac{e^{ik\cdot\left(z-z^{\prime}\right)}}{k^{2}}\label{eq:Long_range_integration}
\end{equation}

The resulting fermionic Lagrangian can now be written exclusively in terms of the degrees of freedom in lower dimension $d+1$: 
\begin{equation}
{\cal L}={\cal L}_M^{d+1}+\frac{e^{2}}{2}\int d^{d+1}x^{\prime} j_{d}^{\mu}\left(x\right)j_{d}^{\mu}\left(x^{\prime}\right)
\times \left.G_{D}\left(z\right) \right|_{\begin{array}{ll}
\left(z_{0},\ldots z_{d}\right)=\left(x_{0}-x_{0}^{\prime},\ldots x_{d}-x_{d}^{\prime}\right)\\
z_{d+1},\cdots z_{D+1}=0
\end{array}}\label{eq:Ld}
\end{equation}

One can also represent the LR interaction in an 
operator form which will be useful later. It consists 
in taking the operator 
$\left(-\left.\partial_{\mu}^{2}\right|_{\mu=0,\ldots,D}\right)^{-1}$
and integrate out the extra dimensions keeping the Laplacian for the
lower dimensions. Here we adopt the notation 
$\partial^{2}=\partial_{0}^{2}+\ldots\partial_{d}^{2}$ and represent the 
interaction as $\frac{e^2}{2}j_\mu \hat{G}_{D\rightarrow d}
\left(-\partial^2\right) j_\mu$, where 
\begin{equation}
\hat{G}_{D\rightarrow d}\equiv G_{D\rightarrow d}\left(-\left.\partial_{\mu}^{2}\right|_{\mu=0,\ldots,,d}\right)
=\int\frac{d^{D-d}k}{\left(2\pi\right)^{D-d}}\frac{1}{-\partial_{0}^{2}-\ldots-\partial_{d+1}^{2}+k_{1}^{2}+\ldots+k_{D-d}^{2}}.\label{eq:GDd_operator}
\end{equation}
The two forms of presenting the resulting Lagrangian emphasize two
different aspects. When writing, as in the first case, the interaction
in terms of a space-time function $G_{D}\left(z\right)$, we see that
the current-current interaction does not depend on the lower dimension
and only on the upper dimension. In turn, when writing as above in
terms of a modified dispersion relation, we see that the formal structure
of the function $G_{D\rightarrow d}$, which will have as argument
the Laplacian, only depends on how many dimensions we are integrating
out. Of course the two approaches are equivalent and, in fact while the function
$G_{D\rightarrow d}$ only depends on the difference between the
dimensions, the operator $\hat{G}_{D\rightarrow d}\equiv G_{D\rightarrow d}\left(-\partial^{2}\right)$
does not. The interplay of the two ways of looking at the theory are
equally useful. 

As mentioned before, it is possible to transfer this
LR interaction into the the kinetic part of gauge fields 
now living in $d+1$ dimensions as well. The goal is to identify the 
effective theory with:
\begin{equation}
{\cal L}_{d}={\cal L}^{d+1}_M-ie j^{\mu}_{d+1}A_{\mu}+\frac{1}{4}F_{\mu\nu}\hat{M}_{D\rightarrow d}F_{\mu\nu}\label{eq:LongRange_GaugeFields}
\end{equation}
where, unlike Eq.~\eqref{eq:QEDLagrangian}, 
all the fields live in $d+1$ dimensions and the operator $\hat{M}$
is to be fixed in such a way that Eq.\eqref{eq:Ld} is recovered. Note
that this theory is also gauge invariant in lower dimension 
($A_{\mu}\rightarrow A_{\mu}+\frac{1}{e}\partial_{\mu}\alpha$
and $\psi\rightarrow\psi e^{-i\alpha}$). In order to integrate the
gauge fields, a gauge fixing is therefore necessary. Standard gauge
fixing as it was done before is possible, but it is not ideal for our
purposes. A different gauge fixing function which takes into account the
non-locality of the Lagrangian turns out to be more adequate -- 
for more details
see Appendix \ref{sec:Gauge-Fixing}. 
The analogous of the Feynman gauge then cancels the off-diagonal
terms. For this choice, 
the propagator is given by 
$G_{\mu\nu}=\frac{1}{-\partial^{2}}\hat{M}_{D\rightarrow d}^{-1}\delta_{\mu\nu}$. 
By the subsequent integration of the
gauge fields and identification with the Lagrangian (\ref{eq:Ld}) in
the operator form when (\ref{eq:GDd_operator}) is used, 
we conclude that both theories coincide for 
\begin{equation}
\hat{M}_{D\rightarrow d}=\left(-\partial^{2}\hat{G}_{D\rightarrow d}\right)^{-1},\label{eq:MG_relation}
\end{equation}
which concludes the mapping for the modified gauge theory exclusively in $d+1$
dimensions.

In the following we denote the dimensional reduction from a $D+1$ theory 
to a $d+1$ one by the shortcut notation 
``\texorpdfstring{$D\rightarrow d$}{Lg}''.

To conclude this Section we observe that, although the previous treatment 
starts from the relativistic Lagrangian (\ref{eq:QEDLagrangian}), one can as 
well perform in a similar way the dimensional reduction for a non-relativistc 
system coupled to a gauge fields leaving in higher dimensions.

\section{\texorpdfstring{$2\rightarrow1$}{Lg} dimensional reduction: 
construction of dispersion relations\label{sec:Exploring_construction}}

In this Section we consider the $d=1$, $D=2$ case, i.e., the 
\texorpdfstring{$2\rightarrow1$}{Lg} dimensional reduction. Apart 
from being the simplest example of the general dimensional reduction discussed 
in the previous Section, we have 
two main reasons for focusing in detail on such case: {\it i)} the 
possibility of using bosonization considerably simplifies
the treatment; {\it ii)} when dealing with LR interactions, one realizes 
that the dimensionality of the space on which the elementary constituents 
are defined is not crucial -- not as much as in presence of SR interactions -- 
since varying the type and the range of the LR interactions 
one is effectively changing the dimensionality of the system. As an example 
of the last statement we may consider the $1D$ LR Ising model, where 
passing $\sigma$ from $0$ to $1$ one is spanning the effective 
dimensionality from $4$ (which is the upper critical dimension of the SR 
Ising model) to $1$. Therefore controlling the LR interactions one is equivalent (at least, in the renormalization group sense) to 
controlling the dimensionality of the system. 
Furthermore, due to the form of the operator $\hat{G}_{2\rightarrow1}$
and the possibility of mapping the resulting theory through bosonization
(only available when $d=1$), the class
of LR models that can be addressed is
larger once we introduce extra flavors and integrate them. 

We start by analyzing the simplest case where only one flavor and
a single gauge field is present. Before moving to the specific case
of $2\rightarrow1$, we keep the treatment general considering
the more general case $D\rightarrow1$. In this Section 
we take the matter Lagrangian to be the one of 
free Dirac fermions with a current-current local interaction. 
We do not consider current-current interactions between different 
fermionic flavors since the structure of the analysis does not change. 
However in Appendix \ref{sec:Diagram_mapping} we discuss the inclusion of 
such terms and briefly discuss its consequence.

\subsection{One gauge field and one fermionic flavor with a Thirring term
in \texorpdfstring{$D\rightarrow1$}{Lg}\label{sub:D->1}}

The Lagrangian for the massless case is:
\begin{equation}
{\cal L}=-\bar{\psi}\left(\gamma_{\mu}\partial_{\mu}+ie\gamma_{\mu}A_{\mu}\right)\psi+\frac{g}{2}\left(\bar{\psi}\gamma_{\mu}\psi\right)^{2}+\frac{1}{4}F_{\mu\nu}\hat{M}_{D\rightarrow1}F_{\mu\nu}
\label{eq:LD->1}
\end{equation}
The goal of this Section is to obtain an effective Lagrangian for
the bosonic field resulting from bosonization. 
The procedure is closely related to the bosonization of the Thirring model 
\cite{coleman75thirring} and of the Schwinger model \cite{coleman75schwinger}. 
A possible path involves a Hubbard-Stratonovich transformation to replace the four fermion coupling by introducing a vector field $B_\mu$, which is taken here to be such that 
\begin{equation}
\frac{g}{2}\left(\bar{\psi}\gamma_{\mu}\psi\right)^{2}\rightarrow-ieB_{\mu}\left(\bar{\psi}\gamma_{\mu}\psi\right)+\frac{e^{2}}{2g}B_{\mu}^{2}. 
\end{equation}

Each vector field can, in two dimensions, be parameterized by two scalar 
fields as follows: 
$A_{\mu}=\partial_{\mu}\chi-i\varepsilon_{\mu\nu}\partial_{\nu}\varphi$
and $C_{\mu}=\partial_{\mu}\chi^{\prime}-i\varepsilon_{\mu\nu}\partial_{\nu}\varphi^{\prime}$. The coupling between the vector fields and the fermions can 
be eliminated by a phase and chiral transformations. It appears then 
an extra term present due to the chiral anomaly and both $\chi$ and 
$\chi^{\prime}$ decouple from the rest of
the Lagrangian. The new fermionic variables
are then mapped to a free bosonic theory. After a translation of
the bosonic field, the current of the initial fermionic field can be
written in terms of the new bosonic variable $j_{\mu} \equiv 
\pi^{-1/2}\varepsilon_{\mu\nu}\partial_{\nu}\phi$. 
Details of this procedure and of the generalizations discussed below can 
be found in Appendix \ref{sec:Diagram_mapping}.

Finally, integrating the field $\varphi^{\prime}$ the obtained
effective Lagrangian is: 
\begin{equation}
{\cal L}=\frac{1}{2}\left(1+\frac{g}{\pi}\right)\left(\partial_{\mu}\phi\right)^{2}-\frac{e}{\sqrt{\pi}}\partial_{\mu}\phi\partial_{\mu}\varphi-\frac{1}{2}\partial^{2}\varphi\hat{M}_{D\rightarrow1}\partial^{2}\varphi\label{eq:LD->1bosonized}
\end{equation}
The integration of $\varphi$
yields the final Lagrangian:
$${\cal L}=\frac{1}{2}\left(1+\frac{g}{\pi}\right)\left(\partial_{\mu}\phi\right)^{2}+\frac{e^{2}}{2\pi}\phi\hat{M}_{D\rightarrow1}^{-1}\phi.$$
Of course, when $D=1$, then $\hat{M}_{D\rightarrow1}=1$ and the known
result is recovered (see for example \cite{coleman75schwinger} for $g=0$). 

\begin{figure}
\begin{centering}
\includegraphics[scale=0.27]{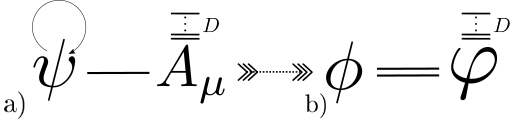}
\par\end{centering}
\protect\caption{Schematic representation of the initial and final theory corresponding to a Pseudo Schwinger-Thirring
model with the gauge field living original in $D+1$ dimensions. The
initial theory, Eq. (\ref{eq:LD->1}), is represented in a) 
having a self interacting fermion (through a current-current interaction)
coupled minimally to a gauge field which originates from an higher
dimensional theory (represented by multiple lines on top of $A_{\mu}$).
In b) we have the resulting theory after bosonization where the fermionic 
degrees of freedom are encoded on $\phi$ and the gauge fields in 
$\varphi$.\label{fig:PseudoST}}
\end{figure}

Since we wish to explore more complicated
theories, it is useful to introduce a diagrammatic representation of
the theories we are working on. In Fig. \ref{fig:PseudoST}, we represent the different fields on
the theory and connect them, if they are coupled to each other, by
straight lines. Straight lines connecting fermionic flavors (including
self coupling) correspond to current-current interaction, lines connecting fermions
to vector fields represent the standard minimal coupling and, finally,
bosons are connected by as many straight lines as there are derivatives
present in their coupling. In the case of the vector fields, we put
as many bars on top of the field as the original dimension that the
field lives in. This means that if the kinetic term is 
$F_{\mu\nu}\hat{M}_{D\rightarrow1}F_{\mu\nu}$
there will be $D$ bars on top of the respective vector field. 
The diagrams do not specify the actual value of the
coupling, even though it can be associated with each line making the
diagrams much more heavy. We plot the initial and
final theories (\ref{eq:LD->1}) and (\ref{eq:LD->1bosonized}) 
in Fig. \ref{fig:PseudoST}.
In Appendix \ref{sec:Diagram_mapping} we use the diagrammatic representation
to represent the intermediate mappings that allow us to establish
the relation between the initial and final theory. We do this systematically
for all the theories represented in the main text.

\subsection{Controlling the kinetic term of bosonic theories}\label{sec:controlling}

\begin{figure}
\begin{centering}
\includegraphics[scale=0.27]{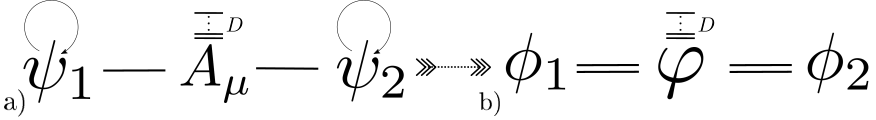}
\par\end{centering}
\protect\caption{Schematic representation of the Pseudo Schwinger-Thirring
model with two flavors and the gauge field living original in $D+1$
dimensions. The initial theory, Eq. (\ref{eq:LD->1}), is represented
in a), while in b) we have the resulting theory after 
bosonization.
\label{fig:PseudoST_2f}}
\end{figure}

The Lagrangians (\ref{eq:LD->1}) and (\ref{eq:LD->1bosonized}) 
are easily generalized to arbitrary number of
flavors $N_{f}$ by introducing a flavor index $\psi\rightarrow\psi_{a}$
, $g\rightarrow g_{a}$ and consequently $\phi\rightarrow\phi_{a}$.
Furthermore, we allow the coupling between the different fermionic
flavors and the gauge field to be different: $e\rightarrow e_{a}$.
In terms of the bosonization procedure this amounts to introduce $N_{f}$
auxiliary fields $B_{\mu}^{a}$ and define a set of new variables 
$C_{\mu}^{a}=A_{\mu}+B_{\mu}^{a}$.
All the rest is analogous to what was described before. Interactions between
different flavors are obtained once the gauge field is integrated out:
${\cal L}=\frac{1}{2}\left(1+\frac{g_{a}}{\pi}\right)\left(\partial_{\mu}\phi_{a}\right)^{2}+\frac{e_{a}e_{b}}{2\pi}\phi_{a}\hat{M}_{D\rightarrow1}^{-1}\phi_{b}$
with implicit sum over flavors. 
Fig. \ref{fig:PseudoST_2f} represents diagramatically
the case of two flavors. Integrating $\phi_{2}$ 
(corresponding to one of the flavors) in the final theory
results in:
\begin{equation}
{\cal L}=\frac{1}{2}\phi_{1}\left[\left(1+\frac{g_{1}}{\pi}\right)\left(-\partial^{2}\right)+\frac{e_{1}^{2}}{\pi}\hat{M}_{D\rightarrow1}^{-1}\right.
\left.-\frac{e_{1}^{2}e_{2}^{2}}{8\pi^{2}}\hat{M}_{D\rightarrow1}^{-1}\frac{1}{\left(1+\frac{g_{2}}{\pi}\right)\left(-\partial^{2}\right)+\frac{e_{2}^{2}}{\pi}\hat{M}_{D\rightarrow1}^{-1}}\hat{M}_{D\rightarrow1}^{-1}\right]\phi_{1}\label{eq:2flavors_effL}
\end{equation}

We now apply the previous results for the case $D=2$. 
From Eq. (\ref{eq:GDd_operator}) 
we see that $\hat{G}_{2\rightarrow1}=\left[2\sqrt{-\partial^{2}}\right]^{-1}$
and therefore $\hat{M}_{D\rightarrow d}^{-1}=\sqrt{-\partial^{2}}/2$ from 
Eq. (\ref{eq:MG_relation}).
For large distances (small momentum) this  is the dominant term of the denominator. 
The relevant scales for this limit can be controlled 
via $g_{2}$ and $e_{2}$. Expanding the denominator one gets
\begin{equation}
\frac{1}{\left(1+\frac{g_{2}}{\pi}\right)\left(-\partial^{2}\right)+\frac{e_{2}^{2}}{\pi}\hat{M}_{2\rightarrow1}^{-1}}
=\frac{\pi}{e_{2}^{2}}\hat{M}_{2\rightarrow1}\underset{n}{\sum}\left[-\left(1+\frac{g_{2}}{\pi}\right)\frac{2\pi}{e_{2}^{2}}\sqrt{-\partial^{2}}\right]^{n}
\end{equation}

Substituting back into the Lagrangian generates a series expansion in 
$\sqrt{-\partial^{2}}$ and therefore
we get all integer and half integer powers $\sqrt{-\partial^{2}}$
in the form:
\begin{equation}
{\cal L}=-\frac{1}{2}\phi\left[\underset{n=1}{\sum}\sigma_{n/2}\left(-\partial^{2}\right)^{\frac{n}{2}}\right]\phi,\label{eq:half_int_expansion}
\end{equation}
where we dropped the index of the remaining field $\phi_1$. 
The first terms of these expansion are 
$\sigma_{1/2}=e_{1}^{2}\left(1-\left(8\pi\right)^{-1}\right)/2\pi$,
$\sigma_{1}=1+\frac{g_{1}}{\pi}+\frac{e_{1}^{2}}{8e_{2}^{2}}
\left(1+\frac{g_{2}}{\pi}\right)$
and for higher terms we have $\sigma_{n/2}=-\frac{e_{1}^{2}}{16\pi}
\left[-\left(1+\frac{g_{2}}{\pi}\right)\frac{2\pi}{e_{2}^{2}}\right]^{n}$.

\begin{figure}
\begin{centering}
\includegraphics[scale=0.27]{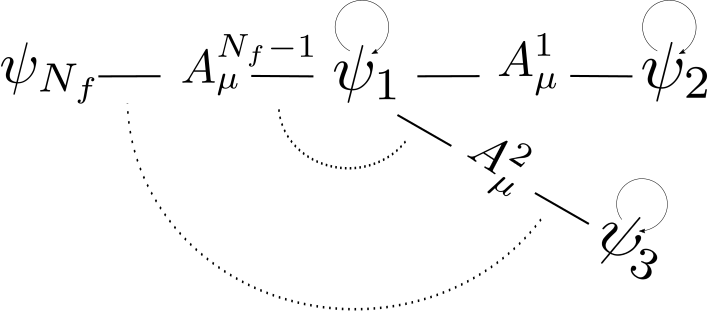}
\par\end{centering}
\protect\caption{Diagramatic representation of the theory with $N_{f}$ 
flavors and
$N_{f}-1$ gauge fields leaving in $2+1$ dimensions. After bosonization
and integration of all gauge fields and all but one fermionic field, 
we can obtain a Lagrangian consisting of a general expansion in integers
and half integer powers of $-\partial^{2}$.\label{fig:general_scheme}}
\end{figure}

There are two main constrains on producing this expansion. For one
side there are only $4$ parameters ($e_{a}$ and $g_{a}$) which
means we cannot control an arbitrary number of terms $\sigma_{n/2}$.
By other side, the sign of $\sigma_{n/2}$ is well defined with $n$
even giving a negative coefficient (except for $n=2$) and
$n$ odd giving positive coefficients. 
One can increase the freedom of choice of the absolute value of the 
coefficients observing that we can enter a third flavor $\psi_{3}$
with a Thirring interaction and a new gauge field $A_{\mu}^{2}$ which
is only coupled to flavors $1$ and $3$. Following the same procedure
of bosonization and integration of the degrees of freedom of the third
flavor, we get a similar expression with new coefficients 
$\sigma_{n/2}^{\mathrm{new}} \equiv \sigma_{n/2}^{(12)}+\sigma_{n/2}^{(13)}$. Here we 
denoted the previous coefficient by $\sigma_{n/2}^{(12)}$ putting
in evidence that it results from an interaction between flavors
$1$ and $2$. Analogously the new contribution is denoted with indices
$(13)$. This procedure can be followed to an arbitrary
number of flavors and one gets an effective coefficient 
$\sigma_{n/2}=\left(1+\frac{g_1}{\pi}\right)\delta_{1,n/2}+\overset{N_{f}}{\underset{i=2}{\sum}}\sigma_{n/2}^{(1i)}$
where $\sigma_{n/2}^{(1i)}=-\frac{e_{1}^{2}}{16\pi}\left[-\left(1+\frac{g_{i}}{\pi}\right)\frac{2\pi}{e_{i}^{2}}\right]^{n}$ for $n\geq1$. The coefficient $\sigma_{1/2}$ does not change and it is controlled exclusively by $e_1$.
By considering more and more
number of flavors in this scheme we are able to control the coefficients
$\sigma_{n/2}$ to an arbitrary order with some constraints. 
The general structure of such theories with $N_{f}$ flavors
is illustrated in Figure \ref{fig:general_scheme}.

\subsection{Controlling fermionic interactions \label{sec:controlling_fermions}}

The integration of the gauge fields naturally leads to the introduction
of non local terms in the fermionic action. In order to obtain the expansion 
on half integer powers (\ref{eq:half_int_expansion}), 
it was crucial to integrate the fermionic degrees of freedom. This enabled
us to overcome the paradigmatic extra quadratic term of
the form $\frac{e^{2}}{2\pi}\phi\hat{M}_{D\rightarrow1}^{-1}\phi$.
Here we apply the same procedure. We want to retain explicitly
one fermionic field while integrating the remaining ones. To this
end we bosonize all but one fermionic degree of freedom. In order
to avoid unnecessary complications, here we restrict ourselves to the
case of two fermionic flavors and no Thirring coupling. Considering
then the two flavors $\psi$ and $\psi^{\prime}$ and bosonizing the
former leads to:
\begin{equation}
{\cal L}=-\bar{\psi}\gamma_{\mu}\partial_{\mu}\psi+ej_{\mu}\varepsilon_{\mu\nu}\partial_{\nu}\varphi-\frac{e^{\prime}}{\sqrt{\pi}}\phi^{\prime}\left(-\partial^{2}\right)\varphi+\frac{1}{2}\left(\partial_{\mu}\phi^{\prime}\right)^{2}
-\frac{1}{2}\varphi\left(-\partial^{2}\right)\hat{M}_{D\rightarrow1}\left(-\partial^{2}\right)\varphi.
\end{equation}
Integrating the degrees of freedom of $\psi^{\prime}$ (in the form
of $\phi^{\prime}$) one gets:
\begin{equation}
{\cal L}=-\bar{\psi}\gamma_{\mu}\partial_{\mu}\psi+ej_{\mu}\varepsilon_{\mu\nu}\partial_{\nu}\varphi-\frac{e_{a}^{2}}{2\pi}\varphi\left(-\partial^{2}\right)\varphi
-\frac{1}{2}\varphi\left(-\partial^{2}\right)\hat{M}_{D\rightarrow1}\left(-\partial^{2}\right)\varphi.
\end{equation}
The final form of the LR fermionic theory is obtained by integrating out
the gauge field, for which it is useful to re-introduce 
$A_{\mu}=-i\varepsilon_{\mu\nu}\partial_{\nu}\varphi$.
The result is:
\begin{equation}
{\cal L}=-\bar{\psi}\gamma_{\mu}\partial_{\mu}\psi+\frac{1}{2}e^{2}j_{\mu}\frac{1}{\frac{e^{\prime2}}{\pi}+\hat{M}_{D\rightarrow1}\left(-\partial^{2}\right)}j_{\mu}.
\end{equation}
Analogously to the case of the bosonic kinetic term, we focus now
on the case $D=2$ and consider the dominant term for large distances.
In this case the denominator is of the form $1+\left(2\pi/e^{\prime2}\right)\sqrt{-\partial^{2}}$. The expansion in Taylor 
series will give then a series of the form
$j_{\mu}\left(-\partial^{2}\right)^{\frac{n}{2}}j_{\mu}$. As in the
previous subsection, the coefficients of the expansion can be chosen with
a certain freedom. This is again done by considering more gauge fields
exactly as in Figure \ref{fig:general_scheme} where we can choose
to include the Thirring terms or not. The final theory obtained is then:
\begin{equation}
{\cal L}=-\bar{\psi}\gamma_{\mu}\partial_{\mu}\psi+\frac{1}{2}e^{2}j_{\mu}\left[\underset{n=1}{\sum}\lambda_{n/2}\left(-\partial^{2}\right)^{\frac{n}{2}}\right]j_{\mu}
\label{eq:fermion_lr}
\end{equation}
where the coefficients $\lambda$ are given by 
\begin{equation}
\lambda_{n/2}=\overset{N_{f}}{\underset{i=2}{\sum}}\frac{\pi}{e_{i}^{2}}\left(-\frac{2\pi}{e_{i}^{2}}\right)^{n}
\end{equation}
in the absence of the Thirring term. 

\section{Experimental implementations \label{exp_implementation}}

In this Section we briefly discuss the recent advances in the simulation 
of gauge fields with cold atoms and trapped ion systems in connection with the formalism discussed 
in Sec.~\ref{sec:fromD_to_d}-\ref{sec:Exploring_construction}.  
As mentioned in Sec.~\ref{intro}, the first experimental demonstration was 
reported last year using trapped ions \cite{Martinez2016}. Several proposals 
have discussed possible implementation schemes of the lattice version of different gauge theories from $1+1$ to $3+1$ dimensions (see Refs.\cite{zohar15,dalmonte16} for a 
review). These proposals either focus on pure gauge systems, or on 
gauge fields coupled to matter, but always with matching dimensions. 
The extension to include a possible mismatch is however straightforward 
since the relevant terms are already present in Hamiltonian and it is only 
needed to suppress fermion hopping in the relevant direction(s). 

As examples we take two different proposals \cite{banerjee12,zohar13} which implement gauge 
symmetry in two different ways using ultracold atoms loaded in optical 
lattices. The dimensional reduction of the space spanned by the fermions 
consists on substituting the periodic lattice potential by a confining 
one in the dimensions to be fixed. This poses no threat to the 
implementation of gauge symmetries. 

More in detail, in the first scheme, gauge symmetry is obtained as a low energy effective symmetry 
by introducing an energy punishment for states that violate gauge symmetry. 
In perturbation theory, the terms that arise are correlated hoppings of bosons 
and fermions, corresponding to matter-gauge coupling, and pure bosonic terms, 
corresponding to the pure gauge contribution for the Hamiltonian. If there 
are no fermions in a given part of the system no correlated hopping would be 
obtained but the pure gauge terms would still be present in perturbation theory.

In the second, gauge symmetry arises from internal symmetries of the system. 
The principle, however, is exactly the same. Due to conservation of total 
hyperfine angular momentum only certain scattering processes are selected. 
With a judicious choice of atomic species only the fermion-boson correlated 
hopping and pure gauge terms will be selected by angular momentum conservation. Again the absence of fermions will retain the former processes corresponding 
exactly to the kind of theories we explore here. Finally, another technique which is applicable to both schemes is to render fermionic tunneling off-resonant in the transverse directions.

\section{Long-range effective Hamiltonians}
\label{LRH}

The Hamiltonians for the effective theories described in the previous
sections are, in general, highly non-trivial. This is the result of non-locality
in time of the Lagrangian. Due to the presence of arbitrarily high powers
of time derivatives, the Euler-Lagrange equations are modified. 
The Hamiltonian formulation of such theories can be achieved within the Ostrogradsky's construction \cite{ostrogradsky50}. The
canonical quantization of theories with non-local kinetic terms, 
like Pseudo QED, was adressed in \cite{barci1997,amaral1992,amorim1999}.

Here, however, we would like to address the canonical quantization of 
fermionic theories presenting non-locality in the interaction term. 
It has been shown that, under certain circumstances, 
and in a perturbative setting, it is possible to use the free equations 
of motion in order to eliminate the non-locality in time 
\cite{knetter1994,arzt1995}. Specifically, such procedure is 
possible when the non-local terms are governed by 
a small coupling parameter. It can then be shown that there 
exists a field transformation equivalent to the application of 
the free equations of motion (by consistently disregarding higher orders of 
the coupling). The fact that the non-locality is obtained from the 
integration of degrees of freedom of a renormalizable theory 
plays a fundamental role \cite{knetter1994}. In that case, 
if there are no unphysical effects at first order approximation 
due to non-locality, we expect that such unphysical effect 
cancel as well at higher orders. This is due to the fact that 
the original theory - that is, the one before integration of degrees of freedom - 
is well defined. Consequently, systematicly disregarding the 
higher powers of the coupling parameter should be consistent 
and the approximation well defined. 
We follow this procedure here in first order perturbation theory.

In previous Sections we have worked with the imaginary time formulation
which is suited to establish the connection with statistical mechanics.
In this Section we construct a quantum Hamiltonian and work in real time.

We focus on the case of non-relativistic fermions, 
as the case of Dirac fermions raises different kinds of questions not to 
be adressed here. In particular, for the case of massless Dirac fermions, 
the free equations of motion imply $\square\psi=0$. Application of 
them in the non-local term will yield, at first order on the coupling, ${\cal O}\left(0\right)$ which is in general divergent. 
Therefore, and according to the previous discussion, perturbation theory 
is not well defined. For non-relativistic fermions this problem is not present 
and, furthermore, in the limit of
large mass we can truncate the temporal derivatives up to some order.
In the following we illustrate this procedure and compute the Hamiltonian
for the lowest order. 

The Lagrangian for the non-relativistic fermion
is then given by:
\begin{equation}
{\cal L}=\psi^{\dagger}\left(i\partial_{0}-eA_{0}+\frac{\hbar^{2}}{2m}\left(\partial_{i}+ieA_{i}\right)^{2}+\mu\right)\psi-\frac{1}{4}F_{\mu\nu}F^{\mu\nu}
\label{non_rel}
\end{equation}
Notice that in Eq. (\ref{non_rel}) there is the extra term proportional to $e^{2}A_{i}^{2}$. This term 
would give rise to
higher order terms: since our treatment is perturbative on $e$, 
those will be discarded.

The current is given by $j_{0}=\psi^{\dagger}\psi$ and $j_{i}=i\frac{\hbar^{2}}{2m}\partial_{i}\psi^{\dagger}\psi$.
The procedure of variable substitution can be done by considering
the current-current interaction in position space: $$j^{\mu}\left(t,x\right){\cal O}\left(t-t^{\prime},x-x^{\prime}\right)j_{\mu}\left(t^{\prime},x^{\prime}\right).$$
Then we consider an expansion of the current of the type 
$$j_{\mu}\left(t^{\prime},x^{\prime}\right)=\sum_{n=0}^{\infty}\frac{\left(t^{\prime}-t\right)^{n}}{n!}\partial_{0}^{n}j_{\mu}\left(t,x^{\prime}\right).$$

Now one can use the equations of motion to replace the time derivatives.
In particular from $\left(i\partial_{0}+\left(\hbar^{2}/2m\right)\partial_{1}^{2}-\mu\right)\psi=0$
one can make the replacement (up to $e^{2}$ order): 
\begin{equation}
\partial_{0}^{n}j_{0}\rightarrow
\overset{n}{\underset{l=0}{\sum}}\left(\begin{array}{c}
n\\
l
\end{array}\right)\left(-i\frac{\hbar^{2}}{2m}\nabla^{2}+i\mu\right)^{l}\psi^{\dagger}\left(i\frac{\hbar^{2}}{2m}\nabla^{2}-i\mu\right)^{2n-l}\psi
\end{equation}
with analogous expressions for the other components of $j_\mu$. 
This term at $x^{\prime}$ is coupled to $j_\mu\left(t,x\right)$ by the function:
\begin{equation} \label{eq:LR_norder}
\int dt^{\prime}{\cal O}\left(t-t^{\prime},x-x^{\prime}\right)\left(t-t^{\prime}\right)^n 
\end{equation}
In this notation ${\cal O}\left(t-t^{\prime},x-x^{\prime}\right)$ are the matrix elements of the operator $G_{3\rightarrow 1}\left(-\partial^{2}\right)$, in its real time form, presented in Appendix \ref{sec:explicit_nonlocal}. 

With such replacements the theory becomes local in time at the cost 
of having (generally complicated and non local) spatial interactions. 
We observe that, in general, since ${\cal O}\left(t,x\right)$ 
is an even function on time, the terms with $n$ odd will not contribute. 

For illustrative purposes we
compute the Hamiltonian density for the case of gauge fields living
in $3+1$ dimensions, chemical potential set to zero and large mass
limit. Each $n$-th derivative above will give rise
to a prefactor $\left(\hbar^{2}/2m\right)^{n}$ and therefore at lowest order
of the large mass limit we can disregard all terms but $n=0$. Furthermore,
as $j_{i}$ is proportional to $\left(\hbar^{2}/2m\right)$, also these terms are of higher order so they are dropped. We are left with the interaction
term:
\begin{equation}
-\frac{e^2}{2} j_{0}\left(t,x\right)\left[\int dt^{\prime}{\cal O}\left(t-t^{\prime},x-x^{\prime}\right)\right]j_{0}\left(t,x^{\prime}\right)
\end{equation}
For the case of gauge fields in $3$ dimensions the effective interaction $\int dt^{\prime}{\cal O}\left(t-t^{\prime},x-x^{\prime}\right)$ is:
\begin{equation}
-\frac{e^2}{8\pi}\int\frac{d^{2}q}{\left(2\pi\right)^{2}}dt^{\prime}\log\left(1+\frac{\Lambda^{2}}{q_{1}^{2}-q_{0}^{2}}\right)e^{-i\left(t-t^{\prime}\right)q_{0}+\left(x-x^{\prime}\right)q_{1}}
\hspace*{-15pt}=-\frac{e^2}{16\pi^{2}\left|x-x^{\prime}\right|}\left(1-e^{-\left|x-x^{\prime}\right|\Lambda}\right)
\end{equation}
Then in the 
limit of large cut-off $\Lambda$ we obtain the effective
Lagrangian:
\begin{equation}
{\cal L}=\psi^{\dagger}\left(t,x\right)\left(i\partial_{0}+\frac{\hbar^{2}}{2m}\partial_{1}^{2}\right)\psi\left(t,x\right)
-\frac{e^{2}}{16\pi^{2}}\int dx^{\prime}\psi^{\dagger}\left(t,x\right)\psi\left(t,x\right)\frac{1}{\left|x-x^{\prime}\right|}\psi^{\dagger}\left(t,x^{\prime}\right)\psi\left(t,x^{\prime}\right)
\end{equation}
This generates an effective Hamiltonian of fermions interacting with
a $1/x$ potential which is the Coulomb potential expected on this
limit. Namely the limit of large massive non-relativistic 
fermions weakly coupled to a three-dimensional gauge field is given by:
\begin{equation}
H=\int dx\left[-\frac{\hbar^{2}}{2m}\psi^{\dagger}\left(t,x\right)\partial_{1}^{2}\psi\left(t,x\right)\right.
\left.+\frac{e^{2}}{16\pi^{2}}\int dx^{\prime}\psi^{\dagger}\left(t,x\right)\psi\left(t,x\right)\frac{1}{\left|x-x^{\prime}\right|}\psi^{\dagger}\left(t,x^{\prime}\right)\psi\left(t,x^{\prime}\right)\right]
\end{equation}

The inclusion of the next leading order on the mass will give rise to 
two new kind of terms: one given by the other current component 
$j_{1}\left(t,x\right)j_{1}\left(t,x^{\prime}\right)/\left|x-x^{\prime}\right|$ 
and the other being a density-density interaction 
coming from Eq. (\ref{eq:LR_norder}) with $n=2$. The later 
will scale as the inverse square of the cut-off and therefore can be dropped in the large cutt-off limit. By other side the the 
first term comming from $j_{1}$ interactions can be interpreted, 
in the lattice language, as a correlated hopping between two fermions at 
a distance $\left|x-x^{\prime}\right|$. This also allows a better understanding of the inital approximation: for large masses the particles 
are slow enough that in lowest order the interaction is simply a 
density-density interaction.

\section{Conclusions}
\label{conclusions}

In this paper we explored a class of models of fermions coupled to gauge fields living in higher dimensions. These models are found to have 
direct application in physical systems, like in graphene, but can have 
a wider range of applicabilities. Here we focused on the possibility 
of mapping long-range (LR) models of statistical mechanics to 
local gauge theories with a dimensional mismatch. Such mappings 
allow one to apply tools that are only available in local theories 
to non-local theories, providing immediate access to insights on the dynamics of the latter. Moreover, the described mapping can be used as a tool to engineer desidered 
LR interactions by a properly engineering the gauge fields in the 
larger dimensionality space(s).

By introducing more fermionic flavors we were able, in the context of 
bosonization, to obtain a kinetic term which consists of an expansion in half integer 
powers of the Laplacian $-\partial^2$. More general expansions in arbitrary 
powers are likely non achievable from this mechanism, since it is 
expected that they would break break unitarity. In fact it was showed 
in Ref.~\cite{Marino14} that the only unitary theories with the pure gauge
 term modified to be 
$\sim F_{\mu\nu}\frac{1}{-\left(\partial^2\right)^\alpha}F_{\mu\nu}$ 
in $2+1$ dimensions are for $\alpha=0$ and $\alpha=1/2$ [notice however 
that (\ref{eq:fermion_lr}) is unitary].  

The coefficients of these expansions display some freedom of 
choice by changing the parameters of the initial local theory. 
They are, however, still bounded by certain conditions,
even though we showed that the freedom of choice can be increased by adding more flavours.  
An interesting question is what kind of non-locality can be 
obtained by a local theory as the ones considered here. 
It would be particularly interesting to investigate 
if the conditions obtained on these coefficients are a consequence of the 
mechanism considered (i.e., a local theory in $d+1$ dimensions with 
minimal coupling between matter and gauge fields) 
and/or if they constitute a physical condition provided by unitarity. 
We also provided an overview on the implementation of this kind of gauge 
theories. In general existing proposals admit a straightforward 
generalization for the realization of artificial gauge theories with dimensional mismatch.

Our procedure can further be generalized by considering additional 
couplings to Higgs fields, interaction between gauge fields or 
other gauge symmetries besides $U\left(1\right)$. The integration 
of bosonic gauge fields or general gauge fields may also enlarge 
the space of LR models obtained after the dimensional reduction and it would be interesting to investigate if one can obtain fermionic interaction expansions 
like (\ref{eq:fermion_lr}) in higher dimensions. 

From the application side, our results can be applicable to a series of 
long-range interacting problems, where a mapping to a local higher dimensional
 field theory would allow to apply generic results for local field theories. 
This includes the characterization of topological order 
(for example towards the extension of the $10$-fold classification to LR 
hopping free fermion theories \cite{Lepori2016}), the spreading of quantum information, and 
the study of localization mechanisms in the presence of LR 
hopping in one-dimensional systems~\cite{Deng:2017aa}. From a different 
perspective, our approach can be potentially applicable to fracton models, 
as the latter, in some specific cases, can be understood as physical 
systems where gauge and matter degrees of freedom effectively live 
in different dimensionality \cite{Pretko:2016aa}.

{\it Acknowledgements:} 
Discussions with N. Defenu, G. Gori, 
C. Morais Smith, G. Palumbo, S. Ruffo, N. Westerberg and U.-J. Wiese 
are gratefully acknowledged. 
The authors acknowledge the hospitality of the Galileo Galilei Institute 
during the workshop 'From Static to Dynamical Gauge Fields with Ultracold 
Atoms', where part of this work was performed.

\bibliography{ArXiv_20170822}

\begin{thebibliography}{87}%
\makeatletter
\providecommand \@ifxundefined [1]{%
 \@ifx{#1\undefined}
}%
\providecommand \@ifnum [1]{%
 \ifnum #1\expandafter \@firstoftwo
 \else \expandafter \@secondoftwo
 \fi
}%
\providecommand \@ifx [1]{%
 \ifx #1\expandafter \@firstoftwo
 \else \expandafter \@secondoftwo
 \fi
}%
\providecommand \natexlab [1]{#1}%
\providecommand \enquote  [1]{``#1''}%
\providecommand \bibnamefont  [1]{#1}%
\providecommand \bibfnamefont [1]{#1}%
\providecommand \citenamefont [1]{#1}%
\providecommand \href@noop [0]{\@secondoftwo}%
\providecommand \href [0]{\begingroup \@sanitize@url \@href}%
\providecommand \@href[1]{\@@startlink{#1}\@@href}%
\providecommand \@@href[1]{\endgroup#1\@@endlink}%
\providecommand \@sanitize@url [0]{\catcode `\\12\catcode `\$12\catcode
  `\&12\catcode `\#12\catcode `\^12\catcode `\_12\catcode `\%12\relax}%
\providecommand \@@startlink[1]{}%
\providecommand \@@endlink[0]{}%
\providecommand \url  [0]{\begingroup\@sanitize@url \@url }%
\providecommand \@url [1]{\endgroup\@href {#1}{\urlprefix }}%
\providecommand \urlprefix  [0]{URL }%
\providecommand \Eprint [0]{\href }%
\providecommand \doibase [0]{http://dx.doi.org/}%
\providecommand \selectlanguage [0]{\@gobble}%
\providecommand \bibinfo  [0]{\@secondoftwo}%
\providecommand \bibfield  [0]{\@secondoftwo}%
\providecommand \translation [1]{[#1]}%
\providecommand \BibitemOpen [0]{}%
\providecommand \bibitemStop [0]{}%
\providecommand \bibitemNoStop [0]{.\EOS\space}%
\providecommand \EOS [0]{\spacefactor3000\relax}%
\providecommand \BibitemShut  [1]{\csname bibitem#1\endcsname}%
\let\auto@bib@innerbib\@empty
\bibitem [{\citenamefont {Bloch}\ \emph {et~al.}(2008)\citenamefont {Bloch},
  \citenamefont {Dalibard},\ and\ \citenamefont {Zwerger}}]{Bloch2008}%
  \BibitemOpen
  \bibfield  {author} {\bibinfo {author} {\bibfnamefont {I.}~\bibnamefont
  {Bloch}}, \bibinfo {author} {\bibfnamefont {J.}~\bibnamefont {Dalibard}}, \
  and\ \bibinfo {author} {\bibfnamefont {W.}~\bibnamefont {Zwerger}},\ }\href
  {http://link.aps.org/doi/10.1103/RevModPhys.80.885} {\bibfield  {journal}
  {\bibinfo  {journal} {Rev. Mod. Phys.}\ }\textbf {\bibinfo {volume} {80}},\
  \bibinfo {pages} {885} (\bibinfo {year} {2008})}\BibitemShut {NoStop}%
\bibitem [{\citenamefont {Lahaye}\ \emph {et~al.}(2009)\citenamefont {Lahaye},
  \citenamefont {Menotti}, \citenamefont {Santos}, \citenamefont {Lewenstein},\
  and\ \citenamefont {Pfau}}]{Lahaye2009}%
  \BibitemOpen
  \bibfield  {author} {\bibinfo {author} {\bibfnamefont {T.}~\bibnamefont
  {Lahaye}}, \bibinfo {author} {\bibfnamefont {C.}~\bibnamefont {Menotti}},
  \bibinfo {author} {\bibfnamefont {L.}~\bibnamefont {Santos}}, \bibinfo
  {author} {\bibfnamefont {M.}~\bibnamefont {Lewenstein}}, \ and\ \bibinfo
  {author} {\bibfnamefont {T.}~\bibnamefont {Pfau}},\ }\href
  {http://stacks.iop.org/0034-4885/72/i=12/a=126401?key=crossref.f3ade09a61129cb9af3a519307f064f7
  http://arxiv.org/abs/0905.0386{\%}5Cnhttp://stacks.iop.org/0034-4885/72/i=12/a=126401?key=crossref.f3ade09a61129cb9af3a519307f064f7{\%}5Cnhttp://arxiv.org/abs/0905.03}
  {\bibfield  {journal} {\bibinfo  {journal} {Reports Prog. Phys.}\ }\textbf
  {\bibinfo {volume} {71}},\ \bibinfo {pages} {126401} (\bibinfo {year}
  {2009})}\BibitemShut {NoStop}%
\bibitem [{\citenamefont {Blatt}\ and\ \citenamefont {Roos}(2012)}]{Blatt2012}%
  \BibitemOpen
  \bibfield  {author} {\bibinfo {author} {\bibfnamefont {R.}~\bibnamefont
  {Blatt}}\ and\ \bibinfo {author} {\bibfnamefont {C.~F.}\ \bibnamefont
  {Roos}},\ }\href {http://www.nature.com/doifinder/10.1038/nphys2252}
  {\bibfield  {journal} {\bibinfo  {journal} {Nat. Phys.}\ }\textbf {\bibinfo
  {volume} {8}},\ \bibinfo {pages} {277} (\bibinfo {year} {2012})}\BibitemShut
  {NoStop}%
\bibitem [{\citenamefont {Saffman}\ \emph {et~al.}(2009)\citenamefont
  {Saffman}, \citenamefont {Walker},\ and\ \citenamefont
  {Molmer}}]{Saffman2010}%
  \BibitemOpen
  \bibfield  {author} {\bibinfo {author} {\bibfnamefont {M.}~\bibnamefont
  {Saffman}}, \bibinfo {author} {\bibfnamefont {T.~G.}\ \bibnamefont {Walker}},
  \ and\ \bibinfo {author} {\bibfnamefont {K.}~\bibnamefont {Molmer}},\ }\href
  {http://link.aps.org/doi/10.1103/RevModPhys.82.2313
  http://arxiv.org/abs/0909.4777 http://dx.doi.org/10.1103/RevModPhys.82.2313}
  {\bibfield  {journal} {\bibinfo  {journal} {Rev. Mod. Phys.}\ }\textbf
  {\bibinfo {volume} {82}},\ \bibinfo {pages} {2313} (\bibinfo {year}
  {2009})}\BibitemShut {NoStop}%
\bibitem [{\citenamefont {Ritsch}\ \emph {et~al.}(2013)\citenamefont {Ritsch},
  \citenamefont {Domokos}, \citenamefont {Brennecke},\ and\ \citenamefont
  {Esslinger}}]{Ritsch2013}%
  \BibitemOpen
  \bibfield  {author} {\bibinfo {author} {\bibfnamefont {H.}~\bibnamefont
  {Ritsch}}, \bibinfo {author} {\bibfnamefont {P.}~\bibnamefont {Domokos}},
  \bibinfo {author} {\bibfnamefont {F.}~\bibnamefont {Brennecke}}, \ and\
  \bibinfo {author} {\bibfnamefont {T.}~\bibnamefont {Esslinger}},\ }\href
  {http://link.aps.org/doi/10.1103/RevModPhys.85.553} {\bibfield  {journal}
  {\bibinfo  {journal} {Rev. Mod. Phys.}\ }\textbf {\bibinfo {volume} {85}},\
  \bibinfo {pages} {553} (\bibinfo {year} {2013})}\BibitemShut {NoStop}%
\bibitem [{\citenamefont {Britton}\ \emph {et~al.}(2012)\citenamefont
  {Britton}, \citenamefont {Sawyer}, \citenamefont {Keith}, \citenamefont
  {Wang}, \citenamefont {Freericks}, \citenamefont {Uys}, \citenamefont
  {Biercuk},\ and\ \citenamefont {Bollinger}}]{Britton2012}%
  \BibitemOpen
  \bibfield  {author} {\bibinfo {author} {\bibfnamefont {J.~W.}\ \bibnamefont
  {Britton}}, \bibinfo {author} {\bibfnamefont {B.~C.}\ \bibnamefont {Sawyer}},
  \bibinfo {author} {\bibfnamefont {A.~C.}\ \bibnamefont {Keith}}, \bibinfo
  {author} {\bibfnamefont {C.-C.~J.}\ \bibnamefont {Wang}}, \bibinfo {author}
  {\bibfnamefont {J.~K.}\ \bibnamefont {Freericks}}, \bibinfo {author}
  {\bibfnamefont {H.}~\bibnamefont {Uys}}, \bibinfo {author} {\bibfnamefont
  {M.~J.}\ \bibnamefont {Biercuk}}, \ and\ \bibinfo {author} {\bibfnamefont
  {J.~J.}\ \bibnamefont {Bollinger}},\ }\href
  {http://www.nature.com/doifinder/10.1038/nature10981} {\bibfield  {journal}
  {\bibinfo  {journal} {Nature}\ }\textbf {\bibinfo {volume} {484}},\ \bibinfo
  {pages} {489} (\bibinfo {year} {2012})}\BibitemShut {NoStop}%
\bibitem [{\citenamefont {Schau{\ss}}\ \emph {et~al.}(2012)\citenamefont
  {Schau{\ss}}, \citenamefont {Cheneau}, \citenamefont {Endres}, \citenamefont
  {Fukuhara}, \citenamefont {Hild}, \citenamefont {Omran}, \citenamefont
  {Pohl}, \citenamefont {Gross}, \citenamefont {Kuhr},\ and\ \citenamefont
  {Bloch}}]{Schauss2012}%
  \BibitemOpen
  \bibfield  {author} {\bibinfo {author} {\bibfnamefont {P.}~\bibnamefont
  {Schau{\ss}}}, \bibinfo {author} {\bibfnamefont {M.}~\bibnamefont {Cheneau}},
  \bibinfo {author} {\bibfnamefont {M.}~\bibnamefont {Endres}}, \bibinfo
  {author} {\bibfnamefont {T.}~\bibnamefont {Fukuhara}}, \bibinfo {author}
  {\bibfnamefont {S.}~\bibnamefont {Hild}}, \bibinfo {author} {\bibfnamefont
  {A.}~\bibnamefont {Omran}}, \bibinfo {author} {\bibfnamefont
  {T.}~\bibnamefont {Pohl}}, \bibinfo {author} {\bibfnamefont {C.}~\bibnamefont
  {Gross}}, \bibinfo {author} {\bibfnamefont {S.}~\bibnamefont {Kuhr}}, \ and\
  \bibinfo {author} {\bibfnamefont {I.}~\bibnamefont {Bloch}},\ }\href
  {http://www.nature.com/doifinder/10.1038/nature11596} {\bibfield  {journal}
  {\bibinfo  {journal} {Nature}\ }\textbf {\bibinfo {volume} {491}},\ \bibinfo
  {pages} {87} (\bibinfo {year} {2012})}\BibitemShut {NoStop}%
\bibitem [{\citenamefont {Aikawa}\ \emph {et~al.}(2012)\citenamefont {Aikawa},
  \citenamefont {Frisch}, \citenamefont {Mark}, \citenamefont {Baier},
  \citenamefont {Rietzler}, \citenamefont {Grimm},\ and\ \citenamefont
  {Ferlaino}}]{Aikawa2012}%
  \BibitemOpen
  \bibfield  {author} {\bibinfo {author} {\bibfnamefont {K.}~\bibnamefont
  {Aikawa}}, \bibinfo {author} {\bibfnamefont {A.}~\bibnamefont {Frisch}},
  \bibinfo {author} {\bibfnamefont {M.}~\bibnamefont {Mark}}, \bibinfo {author}
  {\bibfnamefont {S.}~\bibnamefont {Baier}}, \bibinfo {author} {\bibfnamefont
  {A.}~\bibnamefont {Rietzler}}, \bibinfo {author} {\bibfnamefont
  {R.}~\bibnamefont {Grimm}}, \ and\ \bibinfo {author} {\bibfnamefont
  {F.}~\bibnamefont {Ferlaino}},\ }\href
  {http://link.aps.org/doi/10.1103/PhysRevLett.108.210401} {\bibfield
  {journal} {\bibinfo  {journal} {Phys. Rev. Lett.}\ }\textbf {\bibinfo
  {volume} {108}},\ \bibinfo {pages} {210401} (\bibinfo {year}
  {2012})}\BibitemShut {NoStop}%
\bibitem [{\citenamefont {Lu}\ \emph {et~al.}(2012)\citenamefont {Lu},
  \citenamefont {Burdick},\ and\ \citenamefont {Lev}}]{Lu2012}%
  \BibitemOpen
  \bibfield  {author} {\bibinfo {author} {\bibfnamefont {M.}~\bibnamefont
  {Lu}}, \bibinfo {author} {\bibfnamefont {N.~Q.}\ \bibnamefont {Burdick}}, \
  and\ \bibinfo {author} {\bibfnamefont {B.~L.}\ \bibnamefont {Lev}},\ }\href
  {http://link.aps.org/doi/10.1103/PhysRevLett.108.215301} {\bibfield
  {journal} {\bibinfo  {journal} {Phys. Rev. Lett.}\ }\textbf {\bibinfo
  {volume} {108}},\ \bibinfo {pages} {215301} (\bibinfo {year}
  {2012})}\BibitemShut {NoStop}%
\bibitem [{\citenamefont {Yan}\ \emph {et~al.}(2013)\citenamefont {Yan},
  \citenamefont {Moses}, \citenamefont {Gadway}, \citenamefont {Covey},
  \citenamefont {Hazzard}, \citenamefont {Rey}, \citenamefont {Jin},\ and\
  \citenamefont {Ye}}]{Yan2013}%
  \BibitemOpen
  \bibfield  {author} {\bibinfo {author} {\bibfnamefont {B.}~\bibnamefont
  {Yan}}, \bibinfo {author} {\bibfnamefont {S.~A.}\ \bibnamefont {Moses}},
  \bibinfo {author} {\bibfnamefont {B.}~\bibnamefont {Gadway}}, \bibinfo
  {author} {\bibfnamefont {J.~P.}\ \bibnamefont {Covey}}, \bibinfo {author}
  {\bibfnamefont {K.~R.~A.}\ \bibnamefont {Hazzard}}, \bibinfo {author}
  {\bibfnamefont {A.~M.}\ \bibnamefont {Rey}}, \bibinfo {author} {\bibfnamefont
  {D.~S.}\ \bibnamefont {Jin}}, \ and\ \bibinfo {author} {\bibfnamefont
  {J.}~\bibnamefont {Ye}},\ }\href
  {http://www.nature.com/doifinder/10.1038/nature12483} {\bibfield  {journal}
  {\bibinfo  {journal} {Nature}\ }\textbf {\bibinfo {volume} {501}},\ \bibinfo
  {pages} {521} (\bibinfo {year} {2013})}\BibitemShut {NoStop}%
\bibitem [{\citenamefont {Islam}\ \emph {et~al.}(2013)\citenamefont {Islam},
  \citenamefont {Senko}, \citenamefont {Campbell}, \citenamefont {Korenblit},
  \citenamefont {Smith}, \citenamefont {Lee}, \citenamefont {Edwards},
  \citenamefont {Wang}, \citenamefont {Freericks},\ and\ \citenamefont
  {Monroe}}]{Islam2013}%
  \BibitemOpen
  \bibfield  {author} {\bibinfo {author} {\bibfnamefont {R.}~\bibnamefont
  {Islam}}, \bibinfo {author} {\bibfnamefont {C.}~\bibnamefont {Senko}},
  \bibinfo {author} {\bibfnamefont {W.}~\bibnamefont {Campbell}}, \bibinfo
  {author} {\bibfnamefont {S.}~\bibnamefont {Korenblit}}, \bibinfo {author}
  {\bibfnamefont {J.}~\bibnamefont {Smith}}, \bibinfo {author} {\bibfnamefont
  {A.}~\bibnamefont {Lee}}, \bibinfo {author} {\bibfnamefont {E.}~\bibnamefont
  {Edwards}}, \bibinfo {author} {\bibfnamefont {C.-C.}\ \bibnamefont {Wang}},
  \bibinfo {author} {\bibfnamefont {J.}~\bibnamefont {Freericks}}, \ and\
  \bibinfo {author} {\bibfnamefont {C.}~\bibnamefont {Monroe}},\ }\href
  {http://www.sciencemag.org/cgi/doi/10.1126/science.1232296} {\bibfield
  {journal} {\bibinfo  {journal} {Science}\ }\textbf {\bibinfo {volume}
  {340}},\ \bibinfo {pages} {583} (\bibinfo {year} {2013})}\BibitemShut
  {NoStop}%
\bibitem [{\citenamefont {Richerme}\ \emph {et~al.}(2014)\citenamefont
  {Richerme}, \citenamefont {Gong}, \citenamefont {Lee}, \citenamefont {Senko},
  \citenamefont {Smith}, \citenamefont {Foss-Feig}, \citenamefont {Michalakis},
  \citenamefont {Gorshkov},\ and\ \citenamefont {Monroe}}]{Richerme2014}%
  \BibitemOpen
  \bibfield  {author} {\bibinfo {author} {\bibfnamefont {P.}~\bibnamefont
  {Richerme}}, \bibinfo {author} {\bibfnamefont {Z.-X.}\ \bibnamefont {Gong}},
  \bibinfo {author} {\bibfnamefont {A.}~\bibnamefont {Lee}}, \bibinfo {author}
  {\bibfnamefont {C.}~\bibnamefont {Senko}}, \bibinfo {author} {\bibfnamefont
  {J.}~\bibnamefont {Smith}}, \bibinfo {author} {\bibfnamefont
  {M.}~\bibnamefont {Foss-Feig}}, \bibinfo {author} {\bibfnamefont
  {S.}~\bibnamefont {Michalakis}}, \bibinfo {author} {\bibfnamefont {A.~V.}\
  \bibnamefont {Gorshkov}}, \ and\ \bibinfo {author} {\bibfnamefont
  {C.}~\bibnamefont {Monroe}},\ }\href
  {http://www.nature.com/doifinder/10.1038/nature13450
  http://arxiv.org/abs/1401.5088{\%}5Cnhttp://www.nature.com/doifinder/10.1038/nature13450
  http://arxiv.org/abs/1401.5088 http://dx.doi.org/10.1038/nature13450}
  {\bibfield  {journal} {\bibinfo  {journal} {Nature}\ }\textbf {\bibinfo
  {volume} {511}},\ \bibinfo {pages} {198} (\bibinfo {year}
  {2014})}\BibitemShut {NoStop}%
\bibitem [{\citenamefont {Jurcevic}\ \emph {et~al.}(2014)\citenamefont
  {Jurcevic}, \citenamefont {Lanyon}, \citenamefont {Hauke}, \citenamefont
  {Hempel}, \citenamefont {Zoller}, \citenamefont {Blatt},\ and\ \citenamefont
  {Roos}}]{Jurcevic2014}%
  \BibitemOpen
  \bibfield  {author} {\bibinfo {author} {\bibfnamefont {P.}~\bibnamefont
  {Jurcevic}}, \bibinfo {author} {\bibfnamefont {B.~P.}\ \bibnamefont
  {Lanyon}}, \bibinfo {author} {\bibfnamefont {P.}~\bibnamefont {Hauke}},
  \bibinfo {author} {\bibfnamefont {C.}~\bibnamefont {Hempel}}, \bibinfo
  {author} {\bibfnamefont {P.}~\bibnamefont {Zoller}}, \bibinfo {author}
  {\bibfnamefont {R.}~\bibnamefont {Blatt}}, \ and\ \bibinfo {author}
  {\bibfnamefont {C.~F.}\ \bibnamefont {Roos}},\ }\href
  {http://www.ncbi.nlm.nih.gov/pubmed/25008526
  http://www.nature.com/doifinder/10.1038/nature13461} {\bibfield  {journal}
  {\bibinfo  {journal} {Nature}\ }\textbf {\bibinfo {volume} {511}},\ \bibinfo
  {pages} {202} (\bibinfo {year} {2014})}\BibitemShut {NoStop}%
\bibitem [{\citenamefont {Douglas}\ \emph {et~al.}(2015)\citenamefont
  {Douglas}, \citenamefont {Habibian}, \citenamefont {Hung}, \citenamefont
  {Gorshkov}, \citenamefont {Kimble},\ and\ \citenamefont
  {Chang}}]{Douglas2015}%
  \BibitemOpen
  \bibfield  {author} {\bibinfo {author} {\bibfnamefont {J.~S.}\ \bibnamefont
  {Douglas}}, \bibinfo {author} {\bibfnamefont {H.}~\bibnamefont {Habibian}},
  \bibinfo {author} {\bibfnamefont {C.-L.}\ \bibnamefont {Hung}}, \bibinfo
  {author} {\bibfnamefont {A.~V.}\ \bibnamefont {Gorshkov}}, \bibinfo {author}
  {\bibfnamefont {H.~J.}\ \bibnamefont {Kimble}}, \ and\ \bibinfo {author}
  {\bibfnamefont {D.~E.}\ \bibnamefont {Chang}},\ }\href
  {http://www.nature.com/doifinder/10.1038/nphoton.2015.57} {\bibfield
  {journal} {\bibinfo  {journal} {Nat. Photonics}\ }\textbf {\bibinfo {volume}
  {9}},\ \bibinfo {pages} {326} (\bibinfo {year} {2015})}\BibitemShut {NoStop}%
\bibitem [{\citenamefont {Schempp}\ \emph {et~al.}(2015)\citenamefont
  {Schempp}, \citenamefont {G{\"{u}}nter}, \citenamefont {W{\"{u}}ster},
  \citenamefont {Weidem{\"{u}}ller},\ and\ \citenamefont
  {Whitlock}}]{Schempp2015}%
  \BibitemOpen
  \bibfield  {author} {\bibinfo {author} {\bibfnamefont {H.}~\bibnamefont
  {Schempp}}, \bibinfo {author} {\bibfnamefont {G.}~\bibnamefont
  {G{\"{u}}nter}}, \bibinfo {author} {\bibfnamefont {S.}~\bibnamefont
  {W{\"{u}}ster}}, \bibinfo {author} {\bibfnamefont {M.}~\bibnamefont
  {Weidem{\"{u}}ller}}, \ and\ \bibinfo {author} {\bibfnamefont
  {S.}~\bibnamefont {Whitlock}},\ }\href
  {http://link.aps.org/doi/10.1103/PhysRevLett.115.093002
  http://arxiv.org/abs/1504.01892
  http://dx.doi.org/10.1103/PhysRevLett.115.093002} {\bibfield  {journal}
  {\bibinfo  {journal} {Phys. Rev. Lett.}\ }\textbf {\bibinfo {volume} {115}},\
  \bibinfo {pages} {093002} (\bibinfo {year} {2015})}\BibitemShut {NoStop}%
\bibitem [{\citenamefont {Landig}\ \emph {et~al.}(2015)\citenamefont {Landig},
  \citenamefont {Brennecke}, \citenamefont {Mottl}, \citenamefont {Donner},\
  and\ \citenamefont {Esslinger}}]{Landig2015}%
  \BibitemOpen
  \bibfield  {author} {\bibinfo {author} {\bibfnamefont {R.}~\bibnamefont
  {Landig}}, \bibinfo {author} {\bibfnamefont {F.}~\bibnamefont {Brennecke}},
  \bibinfo {author} {\bibfnamefont {R.}~\bibnamefont {Mottl}}, \bibinfo
  {author} {\bibfnamefont {T.}~\bibnamefont {Donner}}, \ and\ \bibinfo {author}
  {\bibfnamefont {T.}~\bibnamefont {Esslinger}},\ }\href
  {http://arxiv.org/abs/1503.05565 http://dx.doi.org/10.1038/ncomms8046
  http://www.nature.com/doifinder/10.1038/ncomms8046} {\bibfield  {journal}
  {\bibinfo  {journal} {Nat. Commun.}\ }\textbf {\bibinfo {volume} {6}},\
  \bibinfo {pages} {7046} (\bibinfo {year} {2015})}\BibitemShut {NoStop}%
\bibitem [{\citenamefont {Landig}\ \emph {et~al.}(2016)\citenamefont {Landig},
  \citenamefont {Hruby}, \citenamefont {Dogra}, \citenamefont {Landini},
  \citenamefont {Mottl}, \citenamefont {Donner},\ and\ \citenamefont
  {Esslinger}}]{Landig2015a}%
  \BibitemOpen
  \bibfield  {author} {\bibinfo {author} {\bibfnamefont {R.}~\bibnamefont
  {Landig}}, \bibinfo {author} {\bibfnamefont {L.}~\bibnamefont {Hruby}},
  \bibinfo {author} {\bibfnamefont {N.}~\bibnamefont {Dogra}}, \bibinfo
  {author} {\bibfnamefont {M.}~\bibnamefont {Landini}}, \bibinfo {author}
  {\bibfnamefont {R.}~\bibnamefont {Mottl}}, \bibinfo {author} {\bibfnamefont
  {T.}~\bibnamefont {Donner}}, \ and\ \bibinfo {author} {\bibfnamefont
  {T.}~\bibnamefont {Esslinger}},\ }\href {http://arxiv.org/abs/1511.00007
  http://dx.doi.org/10.1038/nature17409
  http://www.nature.com/doifinder/10.1038/nature17409} {\bibfield  {journal}
  {\bibinfo  {journal} {Nature}\ }\textbf {\bibinfo {volume} {532}},\ \bibinfo
  {pages} {476} (\bibinfo {year} {2016})}\BibitemShut {NoStop}%
\bibitem [{\citenamefont {Martinez}\ \emph {et~al.}(2016)\citenamefont
  {Martinez}, \citenamefont {Muschik}, \citenamefont {Schindler}, \citenamefont
  {Nigg}, \citenamefont {Erhard}, \citenamefont {Heyl}, \citenamefont {Hauke},
  \citenamefont {Dalmonte}, \citenamefont {Monz}, \citenamefont {Zoller},\ and\
  \citenamefont {Blatt}}]{Martinez2016}%
  \BibitemOpen
  \bibfield  {author} {\bibinfo {author} {\bibfnamefont {E.~A.}\ \bibnamefont
  {Martinez}}, \bibinfo {author} {\bibfnamefont {C.~A.}\ \bibnamefont
  {Muschik}}, \bibinfo {author} {\bibfnamefont {P.}~\bibnamefont {Schindler}},
  \bibinfo {author} {\bibfnamefont {D.}~\bibnamefont {Nigg}}, \bibinfo {author}
  {\bibfnamefont {A.}~\bibnamefont {Erhard}}, \bibinfo {author} {\bibfnamefont
  {M.}~\bibnamefont {Heyl}}, \bibinfo {author} {\bibfnamefont {P.}~\bibnamefont
  {Hauke}}, \bibinfo {author} {\bibfnamefont {M.}~\bibnamefont {Dalmonte}},
  \bibinfo {author} {\bibfnamefont {T.}~\bibnamefont {Monz}}, \bibinfo {author}
  {\bibfnamefont {P.}~\bibnamefont {Zoller}}, \ and\ \bibinfo {author}
  {\bibfnamefont {R.}~\bibnamefont {Blatt}},\ }\href
  {http://www.ncbi.nlm.nih.gov/pubmed/27337339
  http://www.nature.com/doifinder/10.1038/nature18318} {\bibfield  {journal}
  {\bibinfo  {journal} {Nature}\ }\textbf {\bibinfo {volume} {534}},\ \bibinfo
  {pages} {516} (\bibinfo {year} {2016})}\BibitemShut {NoStop}%
\bibitem [{\citenamefont {Laflorencie}\ \emph {et~al.}(2005)\citenamefont
  {Laflorencie}, \citenamefont {Affleck},\ and\ \citenamefont
  {Berciu}}]{Laflorencie2005}%
  \BibitemOpen
  \bibfield  {author} {\bibinfo {author} {\bibfnamefont {N.}~\bibnamefont
  {Laflorencie}}, \bibinfo {author} {\bibfnamefont {I.}~\bibnamefont
  {Affleck}}, \ and\ \bibinfo {author} {\bibfnamefont {M.}~\bibnamefont
  {Berciu}},\ }\href {http://arxiv.org/abs/cond-mat/0509390
  http://dx.doi.org/10.1088/1742-5468/2005/12/P12001
  http://stacks.iop.org/1742-5468/2005/i=12/a=P12001?key=crossref.a3345d70da457ea209771f379fe147c9}
  {\bibfield  {journal} {\bibinfo  {journal} {J. Stat. Mech. Theory Exp.}\
  }\textbf {\bibinfo {volume} {2005}},\ \bibinfo {pages} {P12001} (\bibinfo
  {year} {2005})}\BibitemShut {NoStop}%
\bibitem [{\citenamefont {Hastings}\ and\ \citenamefont
  {Koma}(2006)}]{Hastings2005}%
  \BibitemOpen
  \bibfield  {author} {\bibinfo {author} {\bibfnamefont {M.~B.}\ \bibnamefont
  {Hastings}}\ and\ \bibinfo {author} {\bibfnamefont {T.}~\bibnamefont
  {Koma}},\ }\href {http://arxiv.org/abs/math-ph/0507008
  http://dx.doi.org/10.1007/s00220-006-0030-4} {\bibfield  {journal} {\bibinfo
  {journal} {Commun. Math. Phys.}\ }\textbf {\bibinfo {volume} {265}},\
  \bibinfo {pages} {781} (\bibinfo {year} {2006})}\BibitemShut {NoStop}%
\bibitem [{\citenamefont {Koffel}\ \emph {et~al.}(2012)\citenamefont {Koffel},
  \citenamefont {Lewenstein},\ and\ \citenamefont {Tagliacozzo}}]{Koffel2012}%
  \BibitemOpen
  \bibfield  {author} {\bibinfo {author} {\bibfnamefont {T.}~\bibnamefont
  {Koffel}}, \bibinfo {author} {\bibfnamefont {M.}~\bibnamefont {Lewenstein}},
  \ and\ \bibinfo {author} {\bibfnamefont {L.}~\bibnamefont {Tagliacozzo}},\
  }\href {http://link.aps.org/doi/10.1103/PhysRevLett.109.267203} {\bibfield
  {journal} {\bibinfo  {journal} {Phys. Rev. Lett.}\ }\textbf {\bibinfo
  {volume} {109}},\ \bibinfo {pages} {267203} (\bibinfo {year}
  {2012})}\BibitemShut {NoStop}%
\bibitem [{\citenamefont {Schachenmayer}\ \emph {et~al.}(2013)\citenamefont
  {Schachenmayer}, \citenamefont {Lanyon}, \citenamefont {Roos},\ and\
  \citenamefont {Daley}}]{Schachenmayer2013}%
  \BibitemOpen
  \bibfield  {author} {\bibinfo {author} {\bibfnamefont {J.}~\bibnamefont
  {Schachenmayer}}, \bibinfo {author} {\bibfnamefont {B.~P.}\ \bibnamefont
  {Lanyon}}, \bibinfo {author} {\bibfnamefont {C.~F.}\ \bibnamefont {Roos}}, \
  and\ \bibinfo {author} {\bibfnamefont {A.~J.}\ \bibnamefont {Daley}},\ }\href
  {http://link.aps.org/doi/10.1103/PhysRevX.3.031015} {\bibfield  {journal}
  {\bibinfo  {journal} {Phys. Rev. X}\ }\textbf {\bibinfo {volume} {3}},\
  \bibinfo {pages} {031015} (\bibinfo {year} {2013})}\BibitemShut {NoStop}%
\bibitem [{\citenamefont {Eisert}\ \emph {et~al.}(2013)\citenamefont {Eisert},
  \citenamefont {van~den Worm}, \citenamefont {Manmana},\ and\ \citenamefont
  {Kastner}}]{Eisert2013}%
  \BibitemOpen
  \bibfield  {author} {\bibinfo {author} {\bibfnamefont {J.}~\bibnamefont
  {Eisert}}, \bibinfo {author} {\bibfnamefont {M.}~\bibnamefont {van~den
  Worm}}, \bibinfo {author} {\bibfnamefont {S.~R.}\ \bibnamefont {Manmana}}, \
  and\ \bibinfo {author} {\bibfnamefont {M.}~\bibnamefont {Kastner}},\ }\href
  {http://link.aps.org/doi/10.1103/PhysRevLett.111.260401} {\bibfield
  {journal} {\bibinfo  {journal} {Phys. Rev. Lett.}\ }\textbf {\bibinfo
  {volume} {111}},\ \bibinfo {pages} {260401} (\bibinfo {year}
  {2013})}\BibitemShut {NoStop}%
\bibitem [{\citenamefont {Gong}\ \emph {et~al.}(2014)\citenamefont {Gong},
  \citenamefont {Foss-Feig}, \citenamefont {Michalakis},\ and\ \citenamefont
  {Gorshkov}}]{Gong2014}%
  \BibitemOpen
  \bibfield  {author} {\bibinfo {author} {\bibfnamefont {Z.~X.}\ \bibnamefont
  {Gong}}, \bibinfo {author} {\bibfnamefont {M.}~\bibnamefont {Foss-Feig}},
  \bibinfo {author} {\bibfnamefont {S.}~\bibnamefont {Michalakis}}, \ and\
  \bibinfo {author} {\bibfnamefont {A.~V.}\ \bibnamefont {Gorshkov}},\ }\href
  {http://arxiv.org/abs/1401.6174
  http://dx.doi.org/10.1103/PhysRevLett.113.030602} {\bibfield  {journal}
  {\bibinfo  {journal} {Phys. Rev. Lett.}\ }\textbf {\bibinfo {volume} {113}}
  (\bibinfo {year} {2014})}\BibitemShut {NoStop}%
\bibitem [{\citenamefont {Damanik}\ \emph {et~al.}(2014)\citenamefont
  {Damanik}, \citenamefont {Lemm}, \citenamefont {Lukic},\ and\ \citenamefont
  {Yessen}}]{Damanik2014}%
  \BibitemOpen
  \bibfield  {author} {\bibinfo {author} {\bibfnamefont {D.}~\bibnamefont
  {Damanik}}, \bibinfo {author} {\bibfnamefont {M.}~\bibnamefont {Lemm}},
  \bibinfo {author} {\bibfnamefont {M.}~\bibnamefont {Lukic}}, \ and\ \bibinfo
  {author} {\bibfnamefont {W.}~\bibnamefont {Yessen}},\ }\href
  {http://link.aps.org/doi/10.1103/PhysRevLett.113.127202} {\bibfield
  {journal} {\bibinfo  {journal} {Phys. Rev. Lett.}\ }\textbf {\bibinfo
  {volume} {113}},\ \bibinfo {pages} {127202} (\bibinfo {year}
  {2014})}\BibitemShut {NoStop}%
\bibitem [{\citenamefont {Vodola}\ \emph {et~al.}(2014)\citenamefont {Vodola},
  \citenamefont {Lepori}, \citenamefont {Ercolessi}, \citenamefont {Gorshkov},\
  and\ \citenamefont {Pupillo}}]{Vodola2014}%
  \BibitemOpen
  \bibfield  {author} {\bibinfo {author} {\bibfnamefont {D.}~\bibnamefont
  {Vodola}}, \bibinfo {author} {\bibfnamefont {L.}~\bibnamefont {Lepori}},
  \bibinfo {author} {\bibfnamefont {E.}~\bibnamefont {Ercolessi}}, \bibinfo
  {author} {\bibfnamefont {A.~V.}\ \bibnamefont {Gorshkov}}, \ and\ \bibinfo
  {author} {\bibfnamefont {G.}~\bibnamefont {Pupillo}},\ }\href
  {http://arxiv.org/abs/1405.5440
  http://dx.doi.org/10.1103/PhysRevLett.113.156402} {\bibfield  {journal}
  {\bibinfo  {journal} {Phys. Rev. Lett.}\ }\textbf {\bibinfo {volume} {113}}
  (\bibinfo {year} {2014})}\BibitemShut {NoStop}%
\bibitem [{\citenamefont {Ares}\ \emph {et~al.}(2015)\citenamefont {Ares},
  \citenamefont {Esteve}, \citenamefont {Falceto},\ and\ \citenamefont {{De
  Queiroz}}}]{Ares2015}%
  \BibitemOpen
  \bibfield  {author} {\bibinfo {author} {\bibfnamefont {F.}~\bibnamefont
  {Ares}}, \bibinfo {author} {\bibfnamefont {J.~G.}\ \bibnamefont {Esteve}},
  \bibinfo {author} {\bibfnamefont {F.}~\bibnamefont {Falceto}}, \ and\
  \bibinfo {author} {\bibfnamefont {A.~R.}\ \bibnamefont {{De Queiroz}}},\
  }\href {https://journals.aps.org/pra/abstract/10.1103/PhysRevA.92.042334}
  {\bibfield  {journal} {\bibinfo  {journal} {Phys. Rev. A}\ }\textbf {\bibinfo
  {volume} {92}} (\bibinfo {year} {2015})}\BibitemShut {NoStop}%
\bibitem [{\citenamefont {Gori}\ \emph {et~al.}(2015)\citenamefont {Gori},
  \citenamefont {Paganelli}, \citenamefont {Sharma}, \citenamefont {Sodano},\
  and\ \citenamefont {Trombettoni}}]{Gori2015}%
  \BibitemOpen
  \bibfield  {author} {\bibinfo {author} {\bibfnamefont {G.}~\bibnamefont
  {Gori}}, \bibinfo {author} {\bibfnamefont {S.}~\bibnamefont {Paganelli}},
  \bibinfo {author} {\bibfnamefont {A.}~\bibnamefont {Sharma}}, \bibinfo
  {author} {\bibfnamefont {P.}~\bibnamefont {Sodano}}, \ and\ \bibinfo {author}
  {\bibfnamefont {A.}~\bibnamefont {Trombettoni}},\ }\href
  {http://link.aps.org/doi/10.1103/PhysRevB.91.245138
  http://arxiv.org/abs/1405.3616 http://dx.doi.org/10.1103/PhysRevB.91.245138}
  {\bibfield  {journal} {\bibinfo  {journal} {Phys. Rev. B}\ }\textbf {\bibinfo
  {volume} {91}},\ \bibinfo {pages} {245138} (\bibinfo {year}
  {2015})}\BibitemShut {NoStop}%
\bibitem [{\citenamefont {Viyuela}\ \emph {et~al.}(2016)\citenamefont
  {Viyuela}, \citenamefont {Vodola}, \citenamefont {Pupillo},\ and\
  \citenamefont {Martin-Delgado}}]{Viyuela2015}%
  \BibitemOpen
  \bibfield  {author} {\bibinfo {author} {\bibfnamefont {O.}~\bibnamefont
  {Viyuela}}, \bibinfo {author} {\bibfnamefont {D.}~\bibnamefont {Vodola}},
  \bibinfo {author} {\bibfnamefont {G.}~\bibnamefont {Pupillo}}, \ and\
  \bibinfo {author} {\bibfnamefont {M.~A.}\ \bibnamefont {Martin-Delgado}},\
  }\href {http://arxiv.org/abs/1511.05018
  http://dx.doi.org/10.1103/PhysRevB.94.125121} {\bibfield  {journal} {\bibinfo
   {journal} {Phys. Rev. B}\ }\textbf {\bibinfo {volume} {94}} (\bibinfo {year}
  {2016})}\BibitemShut {NoStop}%
\bibitem [{\citenamefont {Gong}\ \emph
  {et~al.}(2016{\natexlab{a}})\citenamefont {Gong}, \citenamefont {Maghrebi},
  \citenamefont {Hu}, \citenamefont {Wall}, \citenamefont {Foss-Feig},\ and\
  \citenamefont {Gorshkov}}]{Gong2016}%
  \BibitemOpen
  \bibfield  {author} {\bibinfo {author} {\bibfnamefont {Z.-X.}\ \bibnamefont
  {Gong}}, \bibinfo {author} {\bibfnamefont {M.~F.}\ \bibnamefont {Maghrebi}},
  \bibinfo {author} {\bibfnamefont {A.}~\bibnamefont {Hu}}, \bibinfo {author}
  {\bibfnamefont {M.~L.}\ \bibnamefont {Wall}}, \bibinfo {author}
  {\bibfnamefont {M.}~\bibnamefont {Foss-Feig}}, \ and\ \bibinfo {author}
  {\bibfnamefont {A.~V.}\ \bibnamefont {Gorshkov}},\ }\href
  {http://link.aps.org/doi/10.1103/PhysRevB.93.041102
  http://arxiv.org/abs/1505.03146{\%}5Cnhttp://dx.doi.org/10.1103/PhysRevB.93.041102{\%}5Cnhttp://link.aps.org/doi/10.1103/PhysRevB.93.041102
  http://arxiv.org/abs/1505.03146 http://dx.doi.org/10.1103/PhysRevB.93.}
  {\bibfield  {journal} {\bibinfo  {journal} {Phys. Rev. B}\ }\textbf {\bibinfo
  {volume} {93}},\ \bibinfo {pages} {041102} (\bibinfo {year}
  {2016}{\natexlab{a}})}\BibitemShut {NoStop}%
\bibitem [{\citenamefont {Lepori}\ and\ \citenamefont
  {Dell'Anna}(2017)}]{Lepori2016}%
  \BibitemOpen
  \bibfield  {author} {\bibinfo {author} {\bibfnamefont {L.}~\bibnamefont
  {Lepori}}\ and\ \bibinfo {author} {\bibfnamefont {L.}~\bibnamefont
  {Dell'Anna}},\ }\href
  {http://iopscience.iop.org/article/10.1088/1367-2630/aa84d0} {\bibfield
  {journal} {\bibinfo  {journal} {New J. Phys.}\ } (\bibinfo {year}
  {2017})}\BibitemShut {NoStop}%
\bibitem [{\citenamefont {Lepori}\ \emph
  {et~al.}(2016{\natexlab{a}})\citenamefont {Lepori}, \citenamefont
  {Trombettoni},\ and\ \citenamefont {Vodola}}]{Lepori2016a}%
  \BibitemOpen
  \bibfield  {author} {\bibinfo {author} {\bibfnamefont {L.}~\bibnamefont
  {Lepori}}, \bibinfo {author} {\bibfnamefont {A.}~\bibnamefont {Trombettoni}},
  \ and\ \bibinfo {author} {\bibfnamefont {D.}~\bibnamefont {Vodola}},\ }\href
  {http://arxiv.org/abs/1607.05358
  http://stacks.iop.org/1742-5468/2017/i=3/a=033102?key=crossref.b0532164c0b131eda4cceb9feab85567}
  {\bibfield  {journal} {\bibinfo  {journal} {J. Stat. Mech. Theory Exp.}\
  }\textbf {\bibinfo {volume} {3}},\ \bibinfo {pages} {1} (\bibinfo {year}
  {2016}{\natexlab{a}})}\BibitemShut {NoStop}%
\bibitem [{\citenamefont {Maghrebi}\ \emph {et~al.}(2016)\citenamefont
  {Maghrebi}, \citenamefont {Gong}, \citenamefont {Foss-Feig},\ and\
  \citenamefont {Gorshkov}}]{Maghrebi2016}%
  \BibitemOpen
  \bibfield  {author} {\bibinfo {author} {\bibfnamefont {M.~F.}\ \bibnamefont
  {Maghrebi}}, \bibinfo {author} {\bibfnamefont {Z.-X.}\ \bibnamefont {Gong}},
  \bibinfo {author} {\bibfnamefont {M.}~\bibnamefont {Foss-Feig}}, \ and\
  \bibinfo {author} {\bibfnamefont {A.~V.}\ \bibnamefont {Gorshkov}},\ }\href
  {http://arxiv.org/abs/1508.00906 http://dx.doi.org/10.1103/PhysRevB.93.125128
  http://link.aps.org/doi/10.1103/PhysRevB.93.125128} {\bibfield  {journal}
  {\bibinfo  {journal} {Phys. Rev. B}\ }\textbf {\bibinfo {volume} {93}},\
  \bibinfo {pages} {125128} (\bibinfo {year} {2016})}\BibitemShut {NoStop}%
\bibitem [{\citenamefont {Santos}\ \emph {et~al.}(2016)\citenamefont {Santos},
  \citenamefont {Borgonovi},\ and\ \citenamefont {Celardo}}]{Santos2016}%
  \BibitemOpen
  \bibfield  {author} {\bibinfo {author} {\bibfnamefont {L.~F.}\ \bibnamefont
  {Santos}}, \bibinfo {author} {\bibfnamefont {F.}~\bibnamefont {Borgonovi}}, \
  and\ \bibinfo {author} {\bibfnamefont {G.~L.}\ \bibnamefont {Celardo}},\
  }\href {http://link.aps.org/doi/10.1103/PhysRevLett.116.250402} {\bibfield
  {journal} {\bibinfo  {journal} {Phys. Rev. Lett.}\ }\textbf {\bibinfo
  {volume} {116}},\ \bibinfo {pages} {250402} (\bibinfo {year}
  {2016})}\BibitemShut {NoStop}%
\bibitem [{\citenamefont {Fey}\ and\ \citenamefont {Schmidt}(2016)}]{Fey2016}%
  \BibitemOpen
  \bibfield  {author} {\bibinfo {author} {\bibfnamefont {S.}~\bibnamefont
  {Fey}}\ and\ \bibinfo {author} {\bibfnamefont {K.~P.}\ \bibnamefont
  {Schmidt}},\ }\href {http://link.aps.org/doi/10.1103/PhysRevB.94.075156}
  {\bibfield  {journal} {\bibinfo  {journal} {Phys. Rev. B}\ }\textbf {\bibinfo
  {volume} {94}},\ \bibinfo {pages} {075156} (\bibinfo {year}
  {2016})}\BibitemShut {NoStop}%
\bibitem [{\citenamefont {Gong}\ \emph
  {et~al.}(2016{\natexlab{b}})\citenamefont {Gong}, \citenamefont {Maghrebi},
  \citenamefont {Hu}, \citenamefont {Foss-Feig}, \citenamefont {Richerme},
  \citenamefont {Monroe},\ and\ \citenamefont {Gorshkov}}]{Gong2016a}%
  \BibitemOpen
  \bibfield  {author} {\bibinfo {author} {\bibfnamefont {Z.~X.}\ \bibnamefont
  {Gong}}, \bibinfo {author} {\bibfnamefont {M.~F.}\ \bibnamefont {Maghrebi}},
  \bibinfo {author} {\bibfnamefont {A.}~\bibnamefont {Hu}}, \bibinfo {author}
  {\bibfnamefont {M.}~\bibnamefont {Foss-Feig}}, \bibinfo {author}
  {\bibfnamefont {P.}~\bibnamefont {Richerme}}, \bibinfo {author}
  {\bibfnamefont {C.}~\bibnamefont {Monroe}}, \ and\ \bibinfo {author}
  {\bibfnamefont {A.~V.}\ \bibnamefont {Gorshkov}},\ }\href
  {http://arxiv.org/abs/1510.02108
  http://dx.doi.org/10.1103/PhysRevB.93.205115} {\bibfield  {journal} {\bibinfo
   {journal} {Phys. Rev. B}\ }\textbf {\bibinfo {volume} {93}} (\bibinfo {year}
  {2016}{\natexlab{b}})}\BibitemShut {NoStop}%
\bibitem [{\citenamefont {Kov{\'{a}}cs}\ \emph {et~al.}(2016)\citenamefont
  {Kov{\'{a}}cs}, \citenamefont {Juh{\'{a}}sz},\ and\ \citenamefont
  {Igl{\'{o}}i}}]{Kovacs2016}%
  \BibitemOpen
  \bibfield  {author} {\bibinfo {author} {\bibfnamefont {I.~A.}\ \bibnamefont
  {Kov{\'{a}}cs}}, \bibinfo {author} {\bibfnamefont {R.}~\bibnamefont
  {Juh{\'{a}}sz}}, \ and\ \bibinfo {author} {\bibfnamefont {F.}~\bibnamefont
  {Igl{\'{o}}i}},\ }\href {http://link.aps.org/doi/10.1103/PhysRevB.93.184203}
  {\bibfield  {journal} {\bibinfo  {journal} {Phys. Rev. B}\ }\textbf {\bibinfo
  {volume} {93}},\ \bibinfo {pages} {184203} (\bibinfo {year}
  {2016})}\BibitemShut {NoStop}%
\bibitem [{\citenamefont {Humeniuk}(2016)}]{Humeniuk2016}%
  \BibitemOpen
  \bibfield  {author} {\bibinfo {author} {\bibfnamefont {S.}~\bibnamefont
  {Humeniuk}},\ }\href {http://link.aps.org/doi/10.1103/PhysRevB.93.104412}
  {\bibfield  {journal} {\bibinfo  {journal} {Phys. Rev. B}\ }\textbf {\bibinfo
  {volume} {93}},\ \bibinfo {pages} {104412} (\bibinfo {year}
  {2016})}\BibitemShut {NoStop}%
\bibitem [{\citenamefont {Bermudez}\ \emph {et~al.}(2016)\citenamefont
  {Bermudez}, \citenamefont {Tagliacozzo}, \citenamefont {Sierra},\ and\
  \citenamefont {Richerme}}]{Bermudez2016}%
  \BibitemOpen
  \bibfield  {author} {\bibinfo {author} {\bibfnamefont {A.}~\bibnamefont
  {Bermudez}}, \bibinfo {author} {\bibfnamefont {L.}~\bibnamefont
  {Tagliacozzo}}, \bibinfo {author} {\bibfnamefont {G.}~\bibnamefont {Sierra}},
  \ and\ \bibinfo {author} {\bibfnamefont {P.}~\bibnamefont {Richerme}},\
  }\href {http://arxiv.org/abs/1607.03337
  http://dx.doi.org/10.1103/PhysRevB.95.024431
  http://link.aps.org/doi/10.1103/PhysRevB.95.024431} {\bibfield  {journal}
  {\bibinfo  {journal} {Phys. Rev. B}\ }\textbf {\bibinfo {volume} {95}},\
  \bibinfo {pages} {024431} (\bibinfo {year} {2016})}\BibitemShut {NoStop}%
\bibitem [{\citenamefont {Lepori}\ \emph
  {et~al.}(2016{\natexlab{b}})\citenamefont {Lepori}, \citenamefont {Vodola},
  \citenamefont {Pupillo}, \citenamefont {Gori},\ and\ \citenamefont
  {Trombettoni}}]{Lepori2016b}%
  \BibitemOpen
  \bibfield  {author} {\bibinfo {author} {\bibfnamefont {L.}~\bibnamefont
  {Lepori}}, \bibinfo {author} {\bibfnamefont {D.}~\bibnamefont {Vodola}},
  \bibinfo {author} {\bibfnamefont {G.}~\bibnamefont {Pupillo}}, \bibinfo
  {author} {\bibfnamefont {G.}~\bibnamefont {Gori}}, \ and\ \bibinfo {author}
  {\bibfnamefont {A.}~\bibnamefont {Trombettoni}},\ }\href
  {http://arxiv.org/abs/1511.05544 http://dx.doi.org/10.1016/j.aop.2016.07.026}
  {\bibfield  {journal} {\bibinfo  {journal} {Ann. Phys.}\ }\textbf {\bibinfo
  {volume} {374}},\ \bibinfo {pages} {35} (\bibinfo {year}
  {2016}{\natexlab{b}})}\BibitemShut {NoStop}%
\bibitem [{\citenamefont {Defenu}\ \emph {et~al.}(2017)\citenamefont {Defenu},
  \citenamefont {Trombettoni},\ and\ \citenamefont {Ruffo}}]{Defenu2017}%
  \BibitemOpen
  \bibfield  {author} {\bibinfo {author} {\bibfnamefont {N.}~\bibnamefont
  {Defenu}}, \bibinfo {author} {\bibfnamefont {A.}~\bibnamefont {Trombettoni}},
  \ and\ \bibinfo {author} {\bibfnamefont {S.}~\bibnamefont {Ruffo}},\ }\href
  {http://arxiv.org/abs/1704.00528} {\bibfield  {journal} {\bibinfo  {journal}
  {eprint arXiv:1704.00528}\ } (\bibinfo {year} {2017})}\BibitemShut {NoStop}%
\bibitem [{\citenamefont {Hauke}\ and\ \citenamefont
  {Tagliacozzo}(2013)}]{Hauke2013}%
  \BibitemOpen
  \bibfield  {author} {\bibinfo {author} {\bibfnamefont {P.}~\bibnamefont
  {Hauke}}\ and\ \bibinfo {author} {\bibfnamefont {L.}~\bibnamefont
  {Tagliacozzo}},\ }\href
  {http://link.aps.org/doi/10.1103/PhysRevLett.111.207202} {\bibfield
  {journal} {\bibinfo  {journal} {Phys. Rev. Lett.}\ }\textbf {\bibinfo
  {volume} {111}},\ \bibinfo {pages} {207202} (\bibinfo {year}
  {2013})}\BibitemShut {NoStop}%
\bibitem [{\citenamefont {Foss-Feig}\ \emph {et~al.}(2015)\citenamefont
  {Foss-Feig}, \citenamefont {Gong}, \citenamefont {Clark},\ and\ \citenamefont
  {Gorshkov}}]{Foss-Feig2015}%
  \BibitemOpen
  \bibfield  {author} {\bibinfo {author} {\bibfnamefont {M.}~\bibnamefont
  {Foss-Feig}}, \bibinfo {author} {\bibfnamefont {Z.-X.}\ \bibnamefont {Gong}},
  \bibinfo {author} {\bibfnamefont {C.~W.}\ \bibnamefont {Clark}}, \ and\
  \bibinfo {author} {\bibfnamefont {A.~V.}\ \bibnamefont {Gorshkov}},\ }\href
  {http://arxiv.org/abs/1410.3466
  http://dx.doi.org/10.1103/PhysRevLett.114.157201
  http://link.aps.org/doi/10.1103/PhysRevLett.114.157201} {\bibfield  {journal}
  {\bibinfo  {journal} {Phys. Rev. Lett.}\ }\textbf {\bibinfo {volume} {114}},\
  \bibinfo {pages} {157201} (\bibinfo {year} {2015})}\BibitemShut {NoStop}%
\bibitem [{\citenamefont {Rajabpour}\ and\ \citenamefont
  {Sotiriadis}(2015)}]{Rajabpour2015}%
  \BibitemOpen
  \bibfield  {author} {\bibinfo {author} {\bibfnamefont {M.~A.}\ \bibnamefont
  {Rajabpour}}\ and\ \bibinfo {author} {\bibfnamefont {S.}~\bibnamefont
  {Sotiriadis}},\ }\href {http://link.aps.org/doi/10.1103/PhysRevB.91.045131}
  {\bibfield  {journal} {\bibinfo  {journal} {Phys. Rev. B}\ }\textbf {\bibinfo
  {volume} {91}},\ \bibinfo {pages} {045131} (\bibinfo {year}
  {2015})}\BibitemShut {NoStop}%
\bibitem [{\citenamefont {Cevolani}\ \emph {et~al.}(2015)\citenamefont
  {Cevolani}, \citenamefont {Carleo},\ and\ \citenamefont
  {Sanchez-Palencia}}]{Cevolani2015}%
  \BibitemOpen
  \bibfield  {author} {\bibinfo {author} {\bibfnamefont {L.}~\bibnamefont
  {Cevolani}}, \bibinfo {author} {\bibfnamefont {G.}~\bibnamefont {Carleo}}, \
  and\ \bibinfo {author} {\bibfnamefont {L.}~\bibnamefont {Sanchez-Palencia}},\
  }\href {http://link.aps.org/doi/10.1103/PhysRevA.92.041603} {\bibfield
  {journal} {\bibinfo  {journal} {Phys. Rev. A}\ }\textbf {\bibinfo {volume}
  {92}},\ \bibinfo {pages} {041603} (\bibinfo {year} {2015})}\BibitemShut
  {NoStop}%
\bibitem [{\citenamefont {Kuwahara}(2016)}]{Kuwahara2016}%
  \BibitemOpen
  \bibfield  {author} {\bibinfo {author} {\bibfnamefont {T.}~\bibnamefont
  {Kuwahara}},\ }\href
  {http://stacks.iop.org/1742-5468/2016/i=5/a=053103?key=crossref.b297593544f320e9bc21800efca2f5f2}
  {\bibfield  {journal} {\bibinfo  {journal} {J. Stat. Mech. Theory Exp.}\
  }\textbf {\bibinfo {volume} {2016}},\ \bibinfo {pages} {053103} (\bibinfo
  {year} {2016})}\BibitemShut {NoStop}%
\bibitem [{\citenamefont {{Van Regemortel}}\ \emph {et~al.}(2016)\citenamefont
  {{Van Regemortel}}, \citenamefont {Sels},\ and\ \citenamefont
  {Wouters}}]{VanRegemortel2016}%
  \BibitemOpen
  \bibfield  {author} {\bibinfo {author} {\bibfnamefont {M.}~\bibnamefont {{Van
  Regemortel}}}, \bibinfo {author} {\bibfnamefont {D.}~\bibnamefont {Sels}}, \
  and\ \bibinfo {author} {\bibfnamefont {M.}~\bibnamefont {Wouters}},\ }\href
  {http://link.aps.org/doi/10.1103/PhysRevA.93.032311} {\bibfield  {journal}
  {\bibinfo  {journal} {Phys. Rev. A}\ }\textbf {\bibinfo {volume} {93}},\
  \bibinfo {pages} {032311} (\bibinfo {year} {2016})}\BibitemShut {NoStop}%
\bibitem [{\citenamefont {Buyskikh}\ \emph {et~al.}(2016)\citenamefont
  {Buyskikh}, \citenamefont {Fagotti}, \citenamefont {Schachenmayer},
  \citenamefont {Essler},\ and\ \citenamefont {Daley}}]{Buyskikh2016}%
  \BibitemOpen
  \bibfield  {author} {\bibinfo {author} {\bibfnamefont {A.~S.}\ \bibnamefont
  {Buyskikh}}, \bibinfo {author} {\bibfnamefont {M.}~\bibnamefont {Fagotti}},
  \bibinfo {author} {\bibfnamefont {J.}~\bibnamefont {Schachenmayer}}, \bibinfo
  {author} {\bibfnamefont {F.}~\bibnamefont {Essler}}, \ and\ \bibinfo {author}
  {\bibfnamefont {A.~J.}\ \bibnamefont {Daley}},\ }\href
  {http://link.aps.org/doi/10.1103/PhysRevA.93.053620} {\bibfield  {journal}
  {\bibinfo  {journal} {Phys. Rev. A}\ }\textbf {\bibinfo {volume} {93}},\
  \bibinfo {pages} {053620} (\bibinfo {year} {2016})}\BibitemShut {NoStop}%
\bibitem [{\citenamefont {Bertini}\ \emph {et~al.}(2015)\citenamefont
  {Bertini}, \citenamefont {Essler}, \citenamefont {Groha},\ and\ \citenamefont
  {Robinson}}]{Bertini2015}%
  \BibitemOpen
  \bibfield  {author} {\bibinfo {author} {\bibfnamefont {B.}~\bibnamefont
  {Bertini}}, \bibinfo {author} {\bibfnamefont {F.~H.~L.}\ \bibnamefont
  {Essler}}, \bibinfo {author} {\bibfnamefont {S.}~\bibnamefont {Groha}}, \
  and\ \bibinfo {author} {\bibfnamefont {N.~J.}\ \bibnamefont {Robinson}},\
  }\href {http://link.aps.org/doi/10.1103/PhysRevLett.115.180601
  http://www.ncbi.nlm.nih.gov/pubmed/26565450} {\bibfield  {journal} {\bibinfo
  {journal} {Phys. Rev. Lett.}\ }\textbf {\bibinfo {volume} {115}},\ \bibinfo
  {pages} {180601} (\bibinfo {year} {2015})}\BibitemShut {NoStop}%
\bibitem [{\citenamefont {Brezin}\ \emph {et~al.}(2014)\citenamefont {Brezin},
  \citenamefont {Parisi},\ and\ \citenamefont {Ricci-Tersenghi}}]{Brezin2014}%
  \BibitemOpen
  \bibfield  {author} {\bibinfo {author} {\bibfnamefont {E.}~\bibnamefont
  {Brezin}}, \bibinfo {author} {\bibfnamefont {G.}~\bibnamefont {Parisi}}, \
  and\ \bibinfo {author} {\bibfnamefont {F.}~\bibnamefont {Ricci-Tersenghi}},\
  }\href {http://link.springer.com/10.1007/s10955-014-1081-0} {\bibfield
  {journal} {\bibinfo  {journal} {J. Stat. Phys.}\ }\textbf {\bibinfo {volume}
  {157}},\ \bibinfo {pages} {855} (\bibinfo {year} {2014})}\BibitemShut
  {NoStop}%
\bibitem [{\citenamefont {Angelini}\ \emph {et~al.}(2014)\citenamefont
  {Angelini}, \citenamefont {Parisi},\ and\ \citenamefont
  {Ricci-Tersenghi}}]{Angelini2014}%
  \BibitemOpen
  \bibfield  {author} {\bibinfo {author} {\bibfnamefont {M.~C.}\ \bibnamefont
  {Angelini}}, \bibinfo {author} {\bibfnamefont {G.}~\bibnamefont {Parisi}}, \
  and\ \bibinfo {author} {\bibfnamefont {F.}~\bibnamefont {Ricci-Tersenghi}},\
  }\href {http://link.aps.org/doi/10.1103/PhysRevE.89.062120} {\bibfield
  {journal} {\bibinfo  {journal} {Phys. Rev. E}\ }\textbf {\bibinfo {volume}
  {89}},\ \bibinfo {pages} {062120} (\bibinfo {year} {2014})}\BibitemShut
  {NoStop}%
\bibitem [{\citenamefont {Defenu}\ \emph {et~al.}(2015)\citenamefont {Defenu},
  \citenamefont {Trombettoni},\ and\ \citenamefont {Codello}}]{Defenu2015}%
  \BibitemOpen
  \bibfield  {author} {\bibinfo {author} {\bibfnamefont {N.}~\bibnamefont
  {Defenu}}, \bibinfo {author} {\bibfnamefont {A.}~\bibnamefont {Trombettoni}},
  \ and\ \bibinfo {author} {\bibfnamefont {A.}~\bibnamefont {Codello}},\ }\href
  {http://link.aps.org/doi/10.1103/PhysRevE.92.052113} {\bibfield  {journal}
  {\bibinfo  {journal} {Phys. Rev. E}\ }\textbf {\bibinfo {volume} {92}},\
  \bibinfo {pages} {052113} (\bibinfo {year} {2015})}\BibitemShut {NoStop}%
\bibitem [{\citenamefont {Behan}\ \emph
  {et~al.}(2017{\natexlab{a}})\citenamefont {Behan}, \citenamefont {Rastelli},
  \citenamefont {Rychkov},\ and\ \citenamefont {Zan}}]{Behan2017}%
  \BibitemOpen
  \bibfield  {author} {\bibinfo {author} {\bibfnamefont {C.}~\bibnamefont
  {Behan}}, \bibinfo {author} {\bibfnamefont {L.}~\bibnamefont {Rastelli}},
  \bibinfo {author} {\bibfnamefont {S.}~\bibnamefont {Rychkov}}, \ and\
  \bibinfo {author} {\bibfnamefont {B.}~\bibnamefont {Zan}},\ }\href
  {https://arxiv.org/abs/1703.05325} {\bibfield  {journal} {\bibinfo  {journal}
  {eprint arXiv:1703.05325}\ } (\bibinfo {year}
  {2017}{\natexlab{a}})}\BibitemShut {NoStop}%
\bibitem [{\citenamefont {Behan}\ \emph
  {et~al.}(2017{\natexlab{b}})\citenamefont {Behan}, \citenamefont {Rastelli},
  \citenamefont {Rychkov},\ and\ \citenamefont {Zan}}]{Behan2017a}%
  \BibitemOpen
  \bibfield  {author} {\bibinfo {author} {\bibfnamefont {C.}~\bibnamefont
  {Behan}}, \bibinfo {author} {\bibfnamefont {L.}~\bibnamefont {Rastelli}},
  \bibinfo {author} {\bibfnamefont {S.}~\bibnamefont {Rychkov}}, \ and\
  \bibinfo {author} {\bibfnamefont {B.}~\bibnamefont {Zan}},\ }\href
  {https://journals.aps.org/prl/abstract/10.1103/PhysRevLett.118.241601}
  {\bibfield  {journal} {\bibinfo  {journal} {Phys. Rev. Lett.}\ }\textbf
  {\bibinfo {volume} {118}},\ \bibinfo {pages} {241601} (\bibinfo {year}
  {2017}{\natexlab{b}})}\BibitemShut {NoStop}%
\bibitem [{\citenamefont {Campa}\ \emph {et~al.}(2014)\citenamefont {Campa},
  \citenamefont {Dauxois}, \citenamefont {Fanelli},\ and\ \citenamefont
  {Ruffo}}]{Campa2014}%
  \BibitemOpen
  \bibfield  {author} {\bibinfo {author} {\bibfnamefont {A.}~\bibnamefont
  {Campa}}, \bibinfo {author} {\bibfnamefont {T.}~\bibnamefont {Dauxois}},
  \bibinfo {author} {\bibfnamefont {D.}~\bibnamefont {Fanelli}}, \ and\
  \bibinfo {author} {\bibfnamefont {S.}~\bibnamefont {Ruffo}},\ }\href@noop {}
  {\emph {\bibinfo {title} {{Physics of Long-Range Interacting Systems}}}}\
  (\bibinfo {year} {2014})\ p.\ \bibinfo {pages} {410}\BibitemShut {NoStop}%
\bibitem [{\citenamefont {Uhlenbeck}\ \emph {et~al.}(1963)\citenamefont
  {Uhlenbeck}, \citenamefont {Hemmer},\ and\ \citenamefont {Kac}}]{Kac1963}%
  \BibitemOpen
  \bibfield  {author} {\bibinfo {author} {\bibfnamefont {G.~E.}\ \bibnamefont
  {Uhlenbeck}}, \bibinfo {author} {\bibfnamefont {P.~C.}\ \bibnamefont
  {Hemmer}}, \ and\ \bibinfo {author} {\bibfnamefont {M.}~\bibnamefont {Kac}},\
  }\href
  {http://scitation.aip.org/content/aip/journal/jmp/4/2/10.1063/1.1703946
  http://link.aip.org/link/JMAPAQ/v4/i2/p216/s1{\&}Agg=doi{\%}5Cnhttp://link.aip.org/link/JMAPAQ/v4/i2/p229/s1{\&}Agg=doi}
  {\bibfield  {journal} {\bibinfo  {journal} {J. Math. Phys.}\ }\textbf
  {\bibinfo {volume} {4}},\ \bibinfo {pages} {216} (\bibinfo {year}
  {1963})}\BibitemShut {NoStop}%
\bibitem [{\citenamefont {Sak}(1973)}]{Sak1973}%
  \BibitemOpen
  \bibfield  {author} {\bibinfo {author} {\bibfnamefont {J.}~\bibnamefont
  {Sak}},\ }\href {http://link.aps.org/doi/10.1103/PhysRevB.8.281} {\bibfield
  {journal} {\bibinfo  {journal} {Phys. Rev. B}\ }\textbf {\bibinfo {volume}
  {8}},\ \bibinfo {pages} {281} (\bibinfo {year} {1973})}\BibitemShut {NoStop}%
\bibitem [{\citenamefont {Mermin}\ and\ \citenamefont
  {Wagner}(1966)}]{mermin1966absence}%
  \BibitemOpen
  \bibfield  {author} {\bibinfo {author} {\bibfnamefont {N.~D.}\ \bibnamefont
  {Mermin}}\ and\ \bibinfo {author} {\bibfnamefont {H.}~\bibnamefont
  {Wagner}},\ }\href
  {https://journals.aps.org/prl/abstract/10.1103/PhysRevLett.17.1133}
  {\bibfield  {journal} {\bibinfo  {journal} {Phys. Rev. Lett.}\ }\textbf
  {\bibinfo {volume} {17}},\ \bibinfo {pages} {1133} (\bibinfo {year}
  {1966})}\BibitemShut {NoStop}%
\bibitem [{\citenamefont {Hohenberg}(1967)}]{hohenberg1967existence}%
  \BibitemOpen
  \bibfield  {author} {\bibinfo {author} {\bibfnamefont {P.}~\bibnamefont
  {Hohenberg}},\ }\href
  {https://journals.aps.org/pr/abstract/10.1103/PhysRev.158.383} {\bibfield
  {journal} {\bibinfo  {journal} {Phys. Rev.}\ }\textbf {\bibinfo {volume}
  {158}},\ \bibinfo {pages} {383} (\bibinfo {year} {1967})}\BibitemShut
  {NoStop}%
\bibitem [{\citenamefont {Vershynina}\ and\ \citenamefont
  {Lieb}(2013)}]{vershynina2013lieb}%
  \BibitemOpen
  \bibfield  {author} {\bibinfo {author} {\bibfnamefont {A.}~\bibnamefont
  {Vershynina}}\ and\ \bibinfo {author} {\bibfnamefont {E.~H.}\ \bibnamefont
  {Lieb}},\ }\href {http://www.scholarpedia.org/article/Lieb-Robinson_bounds}
  {\bibfield  {journal} {\bibinfo  {journal} {Scholarpedia}\ }\textbf {\bibinfo
  {volume} {8}},\ \bibinfo {pages} {31267} (\bibinfo {year}
  {2013})}\BibitemShut {NoStop}%
\bibitem [{\citenamefont {Castro~Neto}\ \emph {et~al.}(2009)\citenamefont
  {Castro~Neto}, \citenamefont {Guinea}, \citenamefont {Peres}, \citenamefont
  {Novoselov},\ and\ \citenamefont {Geim}}]{CastroNeto}%
  \BibitemOpen
  \bibfield  {author} {\bibinfo {author} {\bibfnamefont {A.~H.}\ \bibnamefont
  {Castro~Neto}}, \bibinfo {author} {\bibfnamefont {F.}~\bibnamefont {Guinea}},
  \bibinfo {author} {\bibfnamefont {N.~M.~R.}\ \bibnamefont {Peres}}, \bibinfo
  {author} {\bibfnamefont {K.~S.}\ \bibnamefont {Novoselov}}, \ and\ \bibinfo
  {author} {\bibfnamefont {A.~K.}\ \bibnamefont {Geim}},\ }\href
  {https://link.aps.org/doi/10.1103/RevModPhys.81.109} {\bibfield  {journal}
  {\bibinfo  {journal} {Rev. Mod. Phys.}\ }\textbf {\bibinfo {volume} {81}},\
  \bibinfo {pages} {109} (\bibinfo {year} {2009})}\BibitemShut {NoStop}%
\bibitem [{\citenamefont {Marino}(1993)}]{Marino93}%
  \BibitemOpen
  \bibfield  {author} {\bibinfo {author} {\bibfnamefont {E.}~\bibnamefont
  {Marino}},\ }\href
  {http://www.sciencedirect.com/science/article/pii/0550321393903794}
  {\bibfield  {journal} {\bibinfo  {journal} {Nucl. Phys. B}\ }\textbf
  {\bibinfo {volume} {408}},\ \bibinfo {pages} {551} (\bibinfo {year}
  {1993})}\BibitemShut {NoStop}%
\bibitem [{\citenamefont {Marino}\ \emph {et~al.}(2014)\citenamefont {Marino},
  \citenamefont {Nascimento}, \citenamefont {Alves},\ and\ \citenamefont
  {Smith}}]{Marino14}%
  \BibitemOpen
  \bibfield  {author} {\bibinfo {author} {\bibfnamefont {E.}~\bibnamefont
  {Marino}}, \bibinfo {author} {\bibfnamefont {L.~O.}\ \bibnamefont
  {Nascimento}}, \bibinfo {author} {\bibfnamefont {V.~S.}\ \bibnamefont
  {Alves}}, \ and\ \bibinfo {author} {\bibfnamefont {C.~M.}\ \bibnamefont
  {Smith}},\ }\href
  {https://journals.aps.org/prd/abstract/10.1103/PhysRevD.90.105003} {\bibfield
   {journal} {\bibinfo  {journal} {Phys. Rev. D}\ }\textbf {\bibinfo {volume}
  {90}},\ \bibinfo {pages} {105003} (\bibinfo {year} {2014})}\BibitemShut
  {NoStop}%
\bibitem [{\citenamefont {Kotikov}\ and\ \citenamefont
  {Teber}(2014)}]{Kotikov2014}%
  \BibitemOpen
  \bibfield  {author} {\bibinfo {author} {\bibfnamefont {A.~V.}\ \bibnamefont
  {Kotikov}}\ and\ \bibinfo {author} {\bibfnamefont {S.}~\bibnamefont
  {Teber}},\ }\href {https://link.aps.org/doi/10.1103/PhysRevD.89.065038}
  {\bibfield  {journal} {\bibinfo  {journal} {Phys. Rev. D}\ }\textbf {\bibinfo
  {volume} {89}},\ \bibinfo {pages} {065038} (\bibinfo {year}
  {2014})}\BibitemShut {NoStop}%
\bibitem [{\citenamefont {Marino}\ \emph {et~al.}(2015)\citenamefont {Marino},
  \citenamefont {Nascimento}, \citenamefont {Alves},\ and\ \citenamefont
  {Smith}}]{Marino2015}%
  \BibitemOpen
  \bibfield  {author} {\bibinfo {author} {\bibfnamefont {E.~C.}\ \bibnamefont
  {Marino}}, \bibinfo {author} {\bibfnamefont {L.~O.}\ \bibnamefont
  {Nascimento}}, \bibinfo {author} {\bibfnamefont {V.~S.}\ \bibnamefont
  {Alves}}, \ and\ \bibinfo {author} {\bibfnamefont {C.~M.}\ \bibnamefont
  {Smith}},\ }\href {https://link.aps.org/doi/10.1103/PhysRevX.5.011040}
  {\bibfield  {journal} {\bibinfo  {journal} {Phys. Rev. X}\ }\textbf {\bibinfo
  {volume} {5}},\ \bibinfo {pages} {011040} (\bibinfo {year}
  {2015})}\BibitemShut {NoStop}%
\bibitem [{\citenamefont {Nascimento}\ \emph {et~al.}(2015)\citenamefont
  {Nascimento}, \citenamefont {Alves}, \citenamefont {Pe\~na}, \citenamefont
  {Smith},\ and\ \citenamefont {Marino}}]{Nascimento2015}%
  \BibitemOpen
  \bibfield  {author} {\bibinfo {author} {\bibfnamefont {L.~O.}\ \bibnamefont
  {Nascimento}}, \bibinfo {author} {\bibfnamefont {V.~S.}\ \bibnamefont
  {Alves}}, \bibinfo {author} {\bibfnamefont {F.}~\bibnamefont {Pe\~na}},
  \bibinfo {author} {\bibfnamefont {C.~M.}\ \bibnamefont {Smith}}, \ and\
  \bibinfo {author} {\bibfnamefont {E.~C.}\ \bibnamefont {Marino}},\ }\href
  {https://link.aps.org/doi/10.1103/PhysRevD.92.025018} {\bibfield  {journal}
  {\bibinfo  {journal} {Phys. Rev. D}\ }\textbf {\bibinfo {volume} {92}},\
  \bibinfo {pages} {025018} (\bibinfo {year} {2015})}\BibitemShut {NoStop}%
\bibitem [{\citenamefont {Alves}\ \emph {et~al.}(2017)\citenamefont {Alves},
  \citenamefont {Junior}, \citenamefont {Marino},\ and\ \citenamefont
  {Nascimento}}]{Alves2017}%
  \BibitemOpen
  \bibfield  {author} {\bibinfo {author} {\bibfnamefont {V.~S.}\ \bibnamefont
  {Alves}}, \bibinfo {author} {\bibfnamefont {R.~O.}\ \bibnamefont {Junior}},
  \bibinfo {author} {\bibfnamefont {E.}~\bibnamefont {Marino}}, \ and\ \bibinfo
  {author} {\bibfnamefont {L.~O.}\ \bibnamefont {Nascimento}},\ }\href
  {https://journals.aps.org/prd/abstract/10.1103/PhysRevD.96.034005} {\bibfield
   {journal} {\bibinfo  {journal} {Phys. Rev. D}\ }\textbf {\bibinfo {volume}
  {96}},\ \bibinfo {pages} {034005} (\bibinfo {year} {2017})}\BibitemShut
  {NoStop}%
\bibitem [{\citenamefont {Menezes}\ \emph
  {et~al.}(2016{\natexlab{a}})\citenamefont {Menezes}, \citenamefont
  {Morais~Smith},\ and\ \citenamefont {Palumbo}}]{Menezes2017}%
  \BibitemOpen
  \bibfield  {author} {\bibinfo {author} {\bibfnamefont {N.}~\bibnamefont
  {Menezes}}, \bibinfo {author} {\bibfnamefont {C.}~\bibnamefont
  {Morais~Smith}}, \ and\ \bibinfo {author} {\bibfnamefont {G.}~\bibnamefont
  {Palumbo}},\ }\href {http://arxiv.org/abs/1705.03482} {\bibfield  {journal}
  {\bibinfo  {journal} {eprint arXiv:1705.03482}\ } (\bibinfo {year}
  {2016}{\natexlab{a}})}\BibitemShut {NoStop}%
\bibitem [{\citenamefont {Gorbar}\ \emph {et~al.}(2001)\citenamefont {Gorbar},
  \citenamefont {Gusynin},\ and\ \citenamefont {Miransky}}]{Miransky2001}%
  \BibitemOpen
  \bibfield  {author} {\bibinfo {author} {\bibfnamefont {E.~V.}\ \bibnamefont
  {Gorbar}}, \bibinfo {author} {\bibfnamefont {V.~P.}\ \bibnamefont {Gusynin}},
  \ and\ \bibinfo {author} {\bibfnamefont {V.~A.}\ \bibnamefont {Miransky}},\
  }\href {https://link.aps.org/doi/10.1103/PhysRevD.64.105028} {\bibfield
  {journal} {\bibinfo  {journal} {Phys. Rev. D}\ }\textbf {\bibinfo {volume}
  {64}},\ \bibinfo {pages} {105028} (\bibinfo {year} {2001})}\BibitemShut
  {NoStop}%
\bibitem [{\citenamefont {Menezes}\ \emph
  {et~al.}(2016{\natexlab{b}})\citenamefont {Menezes}, \citenamefont
  {Palumbo},\ and\ \citenamefont {Morais~Smith}}]{Menezes2016}%
  \BibitemOpen
  \bibfield  {author} {\bibinfo {author} {\bibfnamefont {N.}~\bibnamefont
  {Menezes}}, \bibinfo {author} {\bibfnamefont {G.}~\bibnamefont {Palumbo}}, \
  and\ \bibinfo {author} {\bibfnamefont {C.}~\bibnamefont {Morais~Smith}},\
  }\href {http://arxiv.org/abs/1609.05577} {\bibfield  {journal} {\bibinfo
  {journal} {eprint arXiv:1609.05577}\ } (\bibinfo {year}
  {2016}{\natexlab{b}})}\BibitemShut {NoStop}%
\bibitem [{\citenamefont {O'Dell}\ \emph {et~al.}(2000)\citenamefont {O'Dell},
  \citenamefont {Giovanazzi}, \citenamefont {Kurizki},\ and\ \citenamefont
  {Akulin}}]{ODell2000}%
  \BibitemOpen
  \bibfield  {author} {\bibinfo {author} {\bibfnamefont {D.}~\bibnamefont
  {O'Dell}}, \bibinfo {author} {\bibfnamefont {S.}~\bibnamefont {Giovanazzi}},
  \bibinfo {author} {\bibfnamefont {G.}~\bibnamefont {Kurizki}}, \ and\
  \bibinfo {author} {\bibfnamefont {V.~M.}\ \bibnamefont {Akulin}},\ }\href
  {https://link.aps.org/doi/10.1103/PhysRevLett.84.5687} {\bibfield  {journal}
  {\bibinfo  {journal} {Phys. Rev. Lett.}\ }\textbf {\bibinfo {volume} {84}},\
  \bibinfo {pages} {5687} (\bibinfo {year} {2000})}\BibitemShut {NoStop}%
\bibitem [{\citenamefont {Wiese}(2013)}]{wiese13}%
  \BibitemOpen
  \bibfield  {author} {\bibinfo {author} {\bibfnamefont {U.-J.}\ \bibnamefont
  {Wiese}},\ }\href
  {http://onlinelibrary.wiley.com/doi/10.1002/andp.201300104/full} {\bibfield
  {journal} {\bibinfo  {journal} {Ann. Phys.}\ }\textbf {\bibinfo {volume}
  {525}},\ \bibinfo {pages} {777} (\bibinfo {year} {2013})}\BibitemShut
  {NoStop}%
\bibitem [{\citenamefont {Zohar}\ \emph {et~al.}(2015)\citenamefont {Zohar},
  \citenamefont {Cirac},\ and\ \citenamefont {Reznik}}]{zohar15}%
  \BibitemOpen
  \bibfield  {author} {\bibinfo {author} {\bibfnamefont {E.}~\bibnamefont
  {Zohar}}, \bibinfo {author} {\bibfnamefont {J.~I.}\ \bibnamefont {Cirac}}, \
  and\ \bibinfo {author} {\bibfnamefont {B.}~\bibnamefont {Reznik}},\ }\href
  {http://iopscience.iop.org/article/10.1088/0034-4885/79/1/014401/meta}
  {\bibfield  {journal} {\bibinfo  {journal} {Rep. Prog. Phys.}\ }\textbf
  {\bibinfo {volume} {79}},\ \bibinfo {pages} {014401} (\bibinfo {year}
  {2015})}\BibitemShut {NoStop}%
\bibitem [{\citenamefont {Dalmonte}\ and\ \citenamefont
  {Montangero}(2016)}]{dalmonte16}%
  \BibitemOpen
  \bibfield  {author} {\bibinfo {author} {\bibfnamefont {M.}~\bibnamefont
  {Dalmonte}}\ and\ \bibinfo {author} {\bibfnamefont {S.}~\bibnamefont
  {Montangero}},\ }\href
  {http://www.tandfonline.com/doi/abs/10.1080/00107514.2016.1151199} {\bibfield
   {journal} {\bibinfo  {journal} {Contemp. Phys.}\ }\textbf {\bibinfo {volume}
  {57}},\ \bibinfo {pages} {388} (\bibinfo {year} {2016})}\BibitemShut
  {NoStop}%
\bibitem [{\citenamefont {Cardy}(1996)}]{cardy1996}%
  \BibitemOpen
  \bibfield  {author} {\bibinfo {author} {\bibfnamefont {J.}~\bibnamefont
  {Cardy}},\ }\href@noop {} {\emph {\bibinfo {title} {Scaling and
  renormalization in statistical physics}}},\ Vol.~\bibinfo {volume} {5}\
  (\bibinfo  {publisher} {Cambridge university press},\ \bibinfo {year}
  {1996})\BibitemShut {NoStop}%
\bibitem [{\citenamefont {Coleman}(1975)}]{coleman75thirring}%
  \BibitemOpen
  \bibfield  {author} {\bibinfo {author} {\bibfnamefont {S.}~\bibnamefont
  {Coleman}},\ }\href
  {https://journals.aps.org/prd/abstract/10.1103/PhysRevD.11.2088} {\bibfield
  {journal} {\bibinfo  {journal} {Phys. Rev. D}\ }\textbf {\bibinfo {volume}
  {11}},\ \bibinfo {pages} {2088} (\bibinfo {year} {1975})}\BibitemShut
  {NoStop}%
\bibitem [{\citenamefont {Coleman}\ \emph {et~al.}(1975)\citenamefont
  {Coleman}, \citenamefont {Jackiw},\ and\ \citenamefont
  {Susskind}}]{coleman75schwinger}%
  \BibitemOpen
  \bibfield  {author} {\bibinfo {author} {\bibfnamefont {S.}~\bibnamefont
  {Coleman}}, \bibinfo {author} {\bibfnamefont {R.}~\bibnamefont {Jackiw}}, \
  and\ \bibinfo {author} {\bibfnamefont {L.}~\bibnamefont {Susskind}},\ }\href
  {http://www.sciencedirect.com/science/article/pii/0003491675902122}
  {\bibfield  {journal} {\bibinfo  {journal} {Ann. Phys.}\ }\textbf {\bibinfo
  {volume} {93}},\ \bibinfo {pages} {267} (\bibinfo {year} {1975})}\BibitemShut
  {NoStop}%
\bibitem [{\citenamefont {Banerjee}\ \emph {et~al.}(2012)\citenamefont
  {Banerjee}, \citenamefont {Dalmonte}, \citenamefont {M{\"u}ller},
  \citenamefont {Rico}, \citenamefont {Stebler}, \citenamefont {Wiese},\ and\
  \citenamefont {Zoller}}]{banerjee12}%
  \BibitemOpen
  \bibfield  {author} {\bibinfo {author} {\bibfnamefont {D.}~\bibnamefont
  {Banerjee}}, \bibinfo {author} {\bibfnamefont {M.}~\bibnamefont {Dalmonte}},
  \bibinfo {author} {\bibfnamefont {M.}~\bibnamefont {M{\"u}ller}}, \bibinfo
  {author} {\bibfnamefont {E.}~\bibnamefont {Rico}}, \bibinfo {author}
  {\bibfnamefont {P.}~\bibnamefont {Stebler}}, \bibinfo {author} {\bibfnamefont
  {U.-J.}\ \bibnamefont {Wiese}}, \ and\ \bibinfo {author} {\bibfnamefont
  {P.}~\bibnamefont {Zoller}},\ }\href
  {https://journals.aps.org/prl/abstract/10.1103/PhysRevLett.109.175302}
  {\bibfield  {journal} {\bibinfo  {journal} {Phys. Rev. Lett.}\ }\textbf
  {\bibinfo {volume} {109}},\ \bibinfo {pages} {175302} (\bibinfo {year}
  {2012})}\BibitemShut {NoStop}%
\bibitem [{\citenamefont {Zohar}\ \emph {et~al.}(2013)\citenamefont {Zohar},
  \citenamefont {Cirac},\ and\ \citenamefont {Reznik}}]{zohar13}%
  \BibitemOpen
  \bibfield  {author} {\bibinfo {author} {\bibfnamefont {E.}~\bibnamefont
  {Zohar}}, \bibinfo {author} {\bibfnamefont {J.~I.}\ \bibnamefont {Cirac}}, \
  and\ \bibinfo {author} {\bibfnamefont {B.}~\bibnamefont {Reznik}},\ }\href
  {https://journals.aps.org/pra/abstract/10.1103/PhysRevA.88.023617} {\bibfield
   {journal} {\bibinfo  {journal} {Phys. Rev. A}\ }\textbf {\bibinfo {volume}
  {88}},\ \bibinfo {pages} {023617} (\bibinfo {year} {2013})}\BibitemShut
  {NoStop}%
\bibitem [{\citenamefont {Ostrogradsky}(1850)}]{ostrogradsky50}%
  \BibitemOpen
  \bibfield  {author} {\bibinfo {author} {\bibfnamefont {M.~V.}\ \bibnamefont
  {Ostrogradsky}},\ }\href@noop {} {\bibfield  {journal} {\bibinfo  {journal}
  {Mem. Acad. St. Petersbourg}\ }\textbf {\bibinfo {volume} {6}},\ \bibinfo
  {pages} {385} (\bibinfo {year} {1850})}\BibitemShut {NoStop}%
\bibitem [{\citenamefont {Barci}\ and\ \citenamefont
  {Oxman}(1997)}]{barci1997}%
  \BibitemOpen
  \bibfield  {author} {\bibinfo {author} {\bibfnamefont {D.}~\bibnamefont
  {Barci}}\ and\ \bibinfo {author} {\bibfnamefont {L.}~\bibnamefont {Oxman}},\
  }\href
  {http://www.worldscientific.com/doi/abs/10.1142/S0217732397000510?journalCode=mpla}
  {\bibfield  {journal} {\bibinfo  {journal} {Mod. Phys. Lett. A}\ }\textbf
  {\bibinfo {volume} {12}},\ \bibinfo {pages} {493} (\bibinfo {year}
  {1997})}\BibitemShut {NoStop}%
\bibitem [{\citenamefont {Do~Amaral}\ and\ \citenamefont
  {Marino}(1992)}]{amaral1992}%
  \BibitemOpen
  \bibfield  {author} {\bibinfo {author} {\bibfnamefont {R.}~\bibnamefont
  {Do~Amaral}}\ and\ \bibinfo {author} {\bibfnamefont {E.}~\bibnamefont
  {Marino}},\ }\href
  {http://iopscience.iop.org/article/10.1088/0305-4470/25/19/026/meta}
  {\bibfield  {journal} {\bibinfo  {journal} {J. Phys. A: Math. Gen.}\ }\textbf
  {\bibinfo {volume} {25}},\ \bibinfo {pages} {5183} (\bibinfo {year}
  {1992})}\BibitemShut {NoStop}%
\bibitem [{\citenamefont {Amorim}\ and\ \citenamefont
  {Barcelos-Neto}(1999)}]{amorim1999}%
  \BibitemOpen
  \bibfield  {author} {\bibinfo {author} {\bibfnamefont {R.}~\bibnamefont
  {Amorim}}\ and\ \bibinfo {author} {\bibfnamefont {J.}~\bibnamefont
  {Barcelos-Neto}},\ }\href {http://aip.scitation.org/doi/abs/10.1063/1.532677}
  {\bibfield  {journal} {\bibinfo  {journal} {J. Math. Phys.}\ }\textbf
  {\bibinfo {volume} {40}},\ \bibinfo {pages} {585} (\bibinfo {year}
  {1999})}\BibitemShut {NoStop}%
\bibitem [{\citenamefont {Knetter}(1994)}]{knetter1994}%
  \BibitemOpen
  \bibfield  {author} {\bibinfo {author} {\bibfnamefont {C.~G.}\ \bibnamefont
  {Knetter}},\ }\href
  {https://journals.aps.org/prd/abstract/10.1103/PhysRevD.49.6709} {\bibfield
  {journal} {\bibinfo  {journal} {Phys. Rev. D}\ }\textbf {\bibinfo {volume}
  {49}},\ \bibinfo {pages} {6709} (\bibinfo {year} {1994})}\BibitemShut
  {NoStop}%
\bibitem [{\citenamefont {Arzt}(1995)}]{arzt1995}%
  \BibitemOpen
  \bibfield  {author} {\bibinfo {author} {\bibfnamefont {C.}~\bibnamefont
  {Arzt}},\ }\href
  {http://www.sciencedirect.com/science/article/pii/037026939401419D}
  {\bibfield  {journal} {\bibinfo  {journal} {Phys. Lett. B}\ }\textbf
  {\bibinfo {volume} {342}},\ \bibinfo {pages} {189} (\bibinfo {year}
  {1995})}\BibitemShut {NoStop}%
\bibitem [{\citenamefont {Deng}\ \emph {et~al.}(2017)\citenamefont {Deng},
  \citenamefont {Kravtsov}, \citenamefont {Shlyapnikov},\ and\ \citenamefont
  {Santos}}]{Deng:2017aa}%
  \BibitemOpen
  \bibfield  {author} {\bibinfo {author} {\bibfnamefont {X.}~\bibnamefont
  {Deng}}, \bibinfo {author} {\bibfnamefont {V.}~\bibnamefont {Kravtsov}},
  \bibinfo {author} {\bibfnamefont {G.}~\bibnamefont {Shlyapnikov}}, \ and\
  \bibinfo {author} {\bibfnamefont {L.}~\bibnamefont {Santos}},\ }\href
  {https://arxiv.org/abs/1706.04088} {\bibfield  {journal} {\bibinfo  {journal}
  {eprint arXiv:1706.04088}\ } (\bibinfo {year} {2017})}\BibitemShut {NoStop}%
\bibitem [{\citenamefont {Pretko}(2017)}]{Pretko:2016aa}%
  \BibitemOpen
  \bibfield  {author} {\bibinfo {author} {\bibfnamefont {M.}~\bibnamefont
  {Pretko}},\ }\href
  {https://journals.aps.org/prb/abstract/10.1103/PhysRevB.96.035119} {\bibfield
   {journal} {\bibinfo  {journal} {Phys. Rev. B}\ }\textbf {\bibinfo {volume}
  {96}},\ \bibinfo {pages} {035119} (\bibinfo {year} {2017})}\BibitemShut
  {NoStop}%
\end{thebibliography}%

\appendix 

\section{Gauge Fixing\label{sec:Gauge-Fixing}}

The Fadeev-Popov method isolates the spurious degrees of freedom by
introducing in the path integral $1=\int D\alpha\delta\left(G\left(A^{\alpha}\right)\right)\frac{\delta G\left(A^{\alpha}\right)}{\delta\alpha}$
where $A^{\alpha}=A_{\mu}+\frac{1}{e}\partial_{\mu}\alpha$. The gauge
fixing that was used in the first integration corresponds to take
$G\left(A^{\alpha}\right)=\partial_{\mu}A_{\mu}-\omega$ and proceed
with an integration over $\omega$ weighted by $e^{-\omega^{2}/2\xi}$.
The convenience of the Feynman gauge, $\xi=1$, lies in the fact that
the off-diagonal terms of the propagator cancel. Since the kinetic
term of this modified gauge theory is affected by $\hat{M}$, the cancellation
of the off diagonal terms require a gauge fixing depending on $\hat{M}$.
By formally choosing $G\left(A^{\alpha}\right)=\left(M^{-1}\right)^{1/2}\partial_{\mu}A_{\mu}-\omega$
and integrating over $\omega$ with the Gaussian weighting function, 
one finds that 
the resulting propagator is given by:
\begin{equation}
G_{\mu\nu}=\left[\frac{1}{-\partial^{2}}\delta_{\mu\nu}+\left(1-\frac{1}{\xi}\right)\frac{\partial^{\mu}\partial^{\nu}}{\left(-\partial^{2}\right)^{2}}\right]\hat{M}_{D\rightarrow d}^{-1}\label{eq:Propagator_LongRange}
\end{equation}

Alternatively one can consider
the same gauge fixing function as before, 
$G\left(A^{\alpha}\right)=\partial_{\mu}A_{\mu}-\omega$, changing the weighting
factor to $e^{-\omega\hat{M}^{-1}\omega/2\xi}$. Both of the
approaches are related by a simple variable transformation.

\section{General procedure for arbitrary number of fields and diagrammatic\label{sec:Diagram_mapping}}

In this Appendix we present the bosonization procedure used
on these theories accompanied by diagrammatic illustrations. 
Even though the diagrams do not replace the calculations
they become useful to understand the structure of the procedure. The
general strategy consists on four steps:
\begin{enumerate}
\item[1)] Eliminate quartic fermion interaction terms by an Hubbard-Stratonovich
transformation. Here we adopt the notation, 
for the fictitious field, of $B_{\mu}^{ab}$ or
$B_{\mu}^{a}$ in the case $a=b$. It is worth noting that 
for the case of coupling between different flavors a decoupling can be achieved 
by replacing the fermionic interacting term between flavors $a$ and $b$ 
as follows: $g_{ab}j_\mu^aj_\mu^b\rightarrow-ieB_{\mu}^{ab}\left(j_{\mu}^{a}+j_{\mu}^{b}\right)+\frac{e^2}{2g_{ab}}\left(B_{\mu}^{ab}\right)^{2}$. 
The scale $e$ can be chosen arbitrarily according to convenience. 
The integration of $B_{\mu}^{ab}$ generates not only 
the correct coupling between different flavors but also self flavor couplings. 
For this reason it is necessary, in general, to introduce another 
field $B_{\mu}^{a}$ in order to compensate this, even when self 
flavor coupling is absent in the original fermionic Lagrangian.

{\it Diagramatically:} Eliminate any line connecting fermionic
flavors (possibly self coupled) and substitute by a vector field
connecting the two flavors $a$ and $b$. In case of self coupling
there is only a line connecting the vector field to the fermion. 

\item[2)] This point is divided in $3$ main parts:

\begin{enumerate}
\item[2a)] Take the vector fields and parameterize them by two bosonic fields: $A^{\bar{a}}_{\mu}=\partial_{\mu}\chi_{\bar{a}}-i\varepsilon_{\mu\nu}\partial_{\nu}\varphi_{\bar{a}}$ 
and $B_{\mu}^{ab}=\partial_{\mu}\chi_{ab}^\prime-i
\varepsilon_{\mu\nu}\partial_{\nu}\varphi_{ab}^\prime$. Note that the indices 
without bars run through the different flavors and with bars through the 
gauge fields: $a,b\in \left\{1,...,N_f\right\}$ and $\bar{a}\in \left\{1,...,N_g\right\}$.
\item[2b)] Do a chiral transformation eliminating 
the remaining couplings between
fermions and bosons. This is given 
by $\psi_{a}=e^{-i\underset{\bar{b}}{\sum}e_{a\bar{b}}\left(\chi_{\bar{b}}+\gamma_{S}\varphi_{\bar{b}}\right)-ie\underset{b}{\sum}\left(\chi_{ab}+\gamma_{S}\varphi_{ab}\right)}\psi_{a}^{\prime}$. Here $\gamma_s=i\gamma_0\gamma_1$. Due to the chiral anomaly the Lagrangian 
acquires some extra terms in the form of ${\cal L}\rightarrow{\cal L}-\frac{1}{2\pi}\underset{a}{\sum}\left(\underset{\bar{b}}{\sum}e_{a\bar{b}}\partial_{\mu}\varphi_{\bar{b}}+\underset{b}{\sum}e\partial_{\mu}\varphi_{ab}\right)^{2}$
\item[2c)] Map the free fermionic theory to the free 
boson theory $\psi^{\prime}_a\rightarrow\phi^{\prime}_a$. Then we transform 
back the bosonic field: $\phi_{a}^{\prime}=\phi_{a}-\frac{1}{\sqrt{\pi}}\underset{a}{\sum}\left(\underset{\bar{b}}{\sum}e_{a\bar{b}}\partial_{\mu}\varphi_{\bar{b}}+\underset{b}{\sum}e\partial_{\mu}\varphi_{ab}\right)^{2}$. This transformation cancels the term originated by the chiral anomaly. It also creates a coupling between the bosonic fields $\phi_{a}$ and the degrees of freedom associated with the vector fields. As in the case of Section \ref{sec:controlling_fermions} one can retain, without bosonizing, the desired fermionic flavors.
\end{enumerate}

{\it Diagramatically:} Replace fermionic variables $\psi_{a}$
by bosonic flavors $\phi_{a}$, and vector field variables by a respective
bosonic field. All coupling lines become double, signaling that all
interactions have the form $\partial_{\mu}\phi\partial_{\mu}\varphi$. 

\item[3)] Integrate the desired fields.

{\it Diagramatically:} Each bosonic variable has the standard
kinetic term, with exception to the one with bars on top (that originates
from the original gauge field in higher dimensions). When one field is integrated out it
is erased from the diagram and it establishes couplings between 
fields that where connected to it in the previous diagram. 
Furthermore, it changes the kinetic term of all the fields 
that were linked to it. Care is needed at this point since 
if the integrated field is one that originates from a fictitious 
vector field, it just renormalizes the original kinetic term. 
For example in Eq. (\ref{eq:LD->1bosonized}) the integration 
of the fictitious field just renormalized the pre-factor of the kinetic term 
$1\rightarrow1+g/\pi$.

\end{enumerate}

We show this process for the two specific cases used in our
calculations. We consider as well an extra case in which 
there is a current-current coupling between fermions. This serves 
to illustrate how the diagrams can be used to quickly get the structure 
of the theory without performing any calculation.

\subsection{One flavor, self coupled, gauge field originating from \texorpdfstring{$D+1$}{Lg}}

\begin{figure}
\centering{}\includegraphics[scale=0.27]{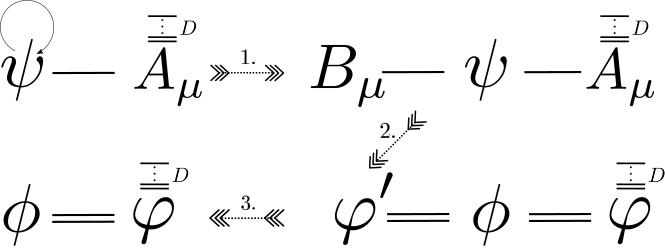}\protect\caption{
Schematic integration of fields from
(\ref{eq:LD->1}) to (\ref{eq:LD->1bosonized}) (see as well 
Figure \ref{fig:PseudoST}).\label{fig:PseudoST_integration}}
\end{figure}

This process is plotted in Figure \ref{fig:PseudoST_integration}.
The numbers on top of the arrows indicate the steps 
described above. In the final diagram where we only
have $\phi$ and $\varphi$ we read immediately that the theory has
the structure:
\begin{equation}
{\cal L}=\frac{\lambda_{1}}{2}\left(\partial_{\mu}\phi\right)^{2}+\lambda_{2}\partial_{\mu}\phi\partial_{\mu}\varphi-\frac{1}{2}\partial^{2}\varphi\hat{M}_{D\rightarrow1}\partial^{2}\varphi.
\end{equation}
The actual values of $\lambda_{1}$ and $\lambda_{2}$ are not obtained
from the diagrams and one has to do the actual computation, getting,  
as in Section \ref{sec:Exploring_construction}, $\lambda_{1}=1+g/\pi$
and $\lambda_{2}=-e/\sqrt{\pi}$, which is Eq. (\ref{eq:LD->1bosonized}) of the 
main text.

\subsection{Two flavors, self coupled, and 
gauge field originating from \texorpdfstring{$D+1$}{Lg}}

\begin{figure}
\centering{}\includegraphics[scale=0.27]{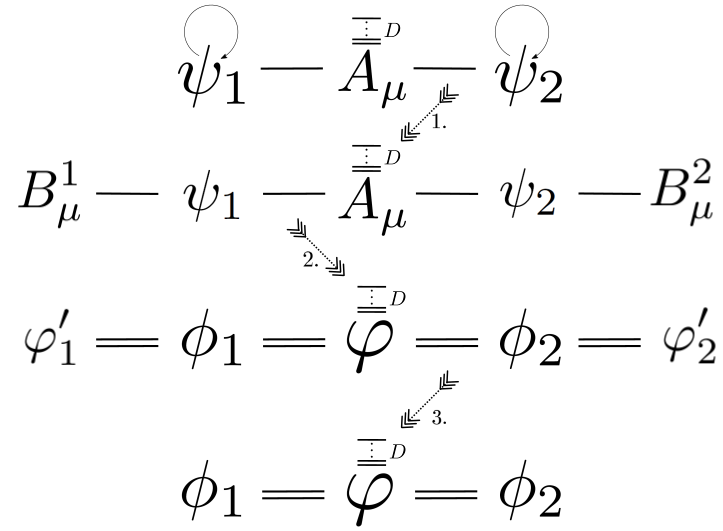}\protect\caption{Schematic integration of fields
in the presence of a gauge field originating from $D+1$ dimensions
interacting with two fermionic flavors.\label{fig:PseudoST_integration_2f}}
\end{figure}

In Figure \ref{fig:PseudoST_integration_2f} we detail the process
of integration of Figure \ref{fig:PseudoST_2f} concerning Sec. \ref{sec:controlling} of the main text. 

\subsection{Two flavors, self coupled, coupled as well to each other and to a gauge field originating from \texorpdfstring{$D+1$}{Lg}}

\begin{figure}
\centering{}\includegraphics[scale=0.27]{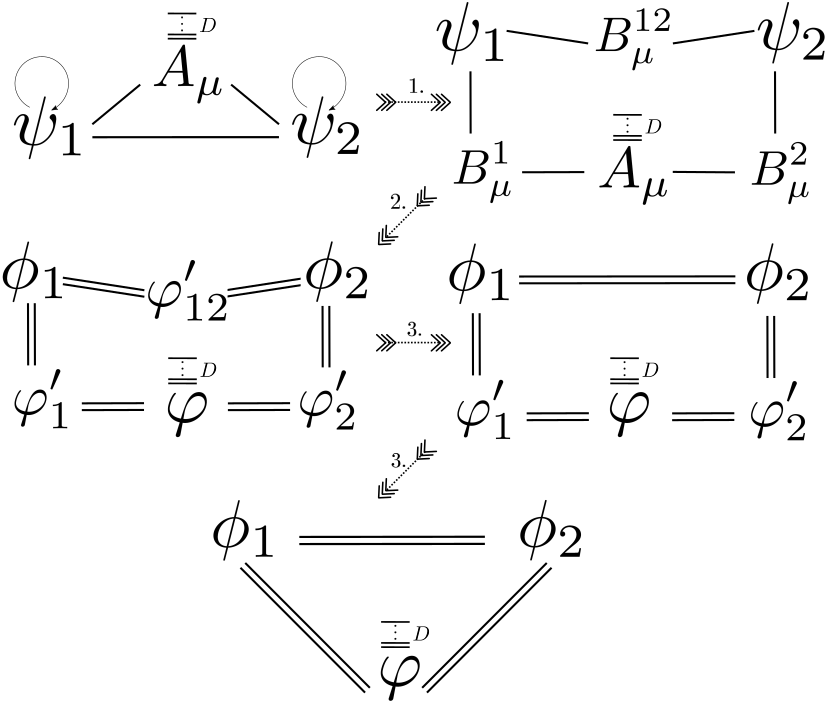}\protect\caption{
Schematic integration of fields
in the presence of a gauge field originating from $D+1$ dimensions
interacting with two fermionic flavors which are self coupled and 
coupled to each other.\label{fig:PseudoST_integration_2f_coupled}}
\end{figure}

The diagrammatic process of considering an initial current-current 
coupling between the fermions is presented in Figure \ref{fig:PseudoST_integration_2f_coupled}.

The resulting theory will have the form:
\begin{equation}
{\cal L}=\frac{\lambda_{1}}{2}\left(\partial_{\mu}\phi_1\right)^{2}+\frac{\lambda_{2}}{2}\left(\partial_{\mu}\phi_2\right)^{2}-\frac{1}{2}\partial^{2}\varphi\hat{M}_{D\rightarrow1}\partial^{2}\varphi
+\lambda_{12}\partial_{\mu}\phi_1\partial_{\mu}\phi_2+\lambda_{1}^{\varphi}\partial_{\mu}\phi_1\partial_{\mu}\varphi
+\lambda_{2}^{\varphi}\partial_{\mu}\phi_2\partial_{\mu}\varphi.
\end{equation}
This new interaction, concerning the inclusion of a current-current interaction between different fermionic flavors, will not change the general expansions \ref{eq:half_int_expansion} or \ref{eq:fermion_lr}, but instead will give an extra freedom on choosing the coefficients.

\section{Non-local quantities for \texorpdfstring{$D$}{Lg} ranging from \texorpdfstring{$1$}{Lg} to \texorpdfstring{$3$}{Lg}} \label{sec:explicit_nonlocal}

The explicit computation for the function (\ref{eq:Long_range_integration})
is possible for the different dimensions. Changing to hyperspherical
coordinates, the integrals can be reduced to $G_{D}\left(z\right)=\frac{\vartheta_{D+1}}{\left(2\pi\right)^{D+1}}\overset{+\infty}{\underset{0}{\int}}dk\overset{\alpha_{D}}{\underset{0}{\int}}d\theta\sin\left(\theta\right)^{D-1}k^{D-2}e^{ik\left|z\right|\cos\theta}$, 
where $\alpha_{D}=\pi$ for any $D$ except $D=1$, where $\alpha_{1}=2\pi$.
For the case of $D=1$ we introduce an IR cut-off $q_{0}$. The results
are given by:
\begin{itemize}
\item $D=1$: 
$$G_{1}\left(z\right)=-\frac{1}{2\pi}\left(\gamma+\frac{1}{2}\log\left(q_{0}\left|z\right|\right)\right)$$
(where $\gamma$ is the Euler's constant and $q_{0}$ the IR cut-off);
\item $D=2$: $$G_{2}\left(z\right)=\frac{1}{4\pi\left|z\right|};$$
\item $D=3$: $$G_{3}\left(z\right)=\frac{1}{4\pi\left|z\right|^{2}}.$$
\end{itemize}
Analogously one can compute the functional form of the various operators
originated from dimensional integration \ref{eq:GDd_operator}. As
explained on the main text, the functional form only depends on the
dimensionality difference $D-d$ while the Laplacian should be the
one of the lower dimensionality $d+1$. We report then the cases $D-d=2$
having in mind $D=3$ and $d=1$ and $D-d=1$ corresponding to
$D=2$ and $d=1$ or $D=3$ and $d=2$. For comparison 
we also write the case trivial case $D-d=0$. The case of $D-d=1$ 
corresponds to Pseudo QED which is already reported in literature 
\cite{Marino93}. We note 
that for $D-d=2$ the integral is divergent and we introduce a UV
cut-off $\Lambda$. This cut-off is for the integrated
dimensions so it can be thought of as a continuous system in $d+1$ dimensions, 
but with a finite lattice spacing in the perpendicular dimensions:
\begin{itemize}
\item $D-d=2$: $$G_{D\rightarrow d}\left(-\partial^{2}\right)=\frac{1}{4\pi}\log\left(\frac{\Lambda^{2}-\partial^{2}}{-\partial^{2}}\right)$$
(where $\Lambda$ is the UV cut-off);
\item $D-d=1$: $$G_{D\rightarrow d}\left(-\partial^{2}\right)=\frac{1}{2}\frac{1}{\sqrt{-\partial^{2}}};$$ 
\item $D=d$: $$G_{D\rightarrow d}\left(-\partial^{2}\right)=\frac{1}{-\partial^{2}}$$
(i.e., the trivial case where no extra dimensions are integrated).
\end{itemize}

\end{document}